\documentclass{jfm}

\usepackage{graphicx}
\usepackage{epsfig,psfrag}
\usepackage{natbib}

\ifCUPmtlplainloaded \else
\checkfont{eurm10}
\iffontfound
\IfFileExists{upmath.sty}
{\typeout{^^JFound AMS Euler Roman fonts on the system,
using the 'upmath' package.^^J}%
\usepackage{upmath}}
{\typeout{^^JFound AMS Euler Roman fonts on the system, but you
dont seem to have the}%
\typeout{'upmath' package installed. JFM.cls can take advantage
of these fonts,^^Jif you use 'upmath' package.^^J}%
}
\else
\fi
\fi
\ifCUPmtlplainloaded \else
\checkfont{msam10}
\iffontfound
\IfFileExists{amssymb.sty}
{\typeout{^^JFound AMS Symbol fonts on the system, using the
'amssymb' package.^^J}%
\usepackage{amssymb}%
 \let\leq=\leqslant
 \let\geq=\geqslant
}{}
\fi
\fi
\ifCUPmtlplainloaded \else
\IfFileExists{amsbsy.sty}
{\typeout{^^JFound the 'amsbsy' package on the system, using it.^^J}%
\usepackage{amsbsy}}
{}
\fi

\newcommand\beq{\begin{equation}}
\newcommand\eeq{\end{equation}}
\newcommand{\half}{\mbox{$\frac12$}}

\title[] {Exchange flow of two immiscible fluids and the principle of
  maximum flux}

\author[R. R. Kerswell]
{R.\ns R.\ns K\ls E\ls R\ls S\ls W\ls E\ls L\ls L${}$
}
\affiliation{School of Mathematics, University of Bristol, University Walk,
Bristol BS8 1TW}

\pubyear{2009}
\volume{???}
\pagerange{???--???}
\date{\today}
\begin{document}

\maketitle

\begin{abstract}

The steady, coaxial flow in which two immiscible, incompressible
fluids move past each other in a cylindrical tube has a continuum of
possibilities due to the arbitrariness of the interface between the
fluids. By invoking the presence of surface tension to at least
restrict the shape of any interface to that of a circular arc or full
circle, we consider the following question: which flow will maximise
the exchange when there is only one dividing interface $\Gamma$?
Surprisingly, the answer differs fundamentally from the better-known
co-directional two-phase flow situation where an axisymmetric
(concentric) core-annular solution always optimises the flux. Instead,
the maximal flux state is invariably asymmetric either being a
`side-by-side' configuration where $\Gamma$ starts and finishes at the
tube wall or an {\em eccentric} core-annular flow where $\Gamma$ is an
off-centre full circle in which the more viscous fluid is surrounded
by the less viscous fluid. The side-by-side solution is the most
efficient exchanger for a small viscosity ratio $\beta \lesssim 4.60$
with an eccentric core-annular solution optimal otherwise. At large
$\beta$, this eccentric solution provides 51\% more flux than the
axisymmetric core-annular flow which is always a local minimiser of
the flux.

\end{abstract}

\section{Introduction}

For Newtonian fluids at least where the governing Navier-Stokes
equations are known, the most fundamental issue in fluid mechanics is
predicting the realised flow solution for a given initial state and
set of boundary conditions against a background of omnipresent noise.
Non-uniqueness of solution is endemic due to the nonlinearity of the
Navier-Stokes equations but even in special limits (e.g. vanishing
Reynolds number or steady, unidirectional flow) where these simplify
to the linear Stokes' equations, degeneracy is rife as
specification of the flow domain is typically part of the problem. A
well-known example of this is the pressure-driven flow of two
immiscible fluids along a cylindrical tube (e.g. Joseph, Renardy \&
Renardy 1984, Joseph, Nguyen and Beavers 1984, and Joseph et
al. 1997). Here there is a continuum of steady unidirectional
solutions possible due to the arbitrariness in the interface between
the two fluids. In practice, however, the axisymmetric core-annular
solution with the more viscous fluid surrounded by the less viscous
fluid is invariably observed for fluid combinations ranging from oil
and water (Charles \& Redberger 1962, Yu \& Sparrow 1967, Hasson, Mann
\& Nir 1970), to molten polymers (Southern \& Ballman 1973, Everage 1973,
Lee \& White 1974, Williams 1975 and Minagawa \& White 1975). 

Interestingly, it appears that if an extra constraint is added to the
system - that the mean volumetric flux along the tube  vanishes -
different steady solutions are observed (Arakeri et al. 2000, Huppert
\& Hallworth 2007, Beckett et al. 2009). Such a flow is easily set up in the
laboratory by placing a tank of dense fluid directly above a tank full
of less dense fluid and connecting the two by a vertical cylindrical
tube. If the density difference or the tube cross-section is small
enough or the fluid viscosities large enough, it is reasonable to
anticipate a steady, coaxial flow established in the tube in which the
denser fluid falls under gravity displacing the less dense fluid
upwards. When the lower tank is initially full and both fluids
incompressible, this exchange flow is constrained to have no net
volume flux along the tube. As in the unidirectional flow situation,
the form of the steady, coaxial two-fluid flow realised is
fascinatingly unclear due to the arbitrariness of the interface
between the fluids (formally, any union of open curves terminating on
the tube wall and closed curves in the interior are possible).  Using
salty and pure water, Arakeri et al (2000) saw only a `half-and-half'
solution where the interface divides the tube cross-section into two
approximately equal domains (hereafter referred to as a
`side-by-side' solution). In contrast, Huppert \& Hallworth (2007) saw
only a concentric core-annular flow as their steady low-Reynolds
solution and recently both types of flow have been seen in the same
apparatus (Beckett et al. 2009).  Beyond its intrinsic interest, this flow
has applications ranging from the exchange of degassed and gas-rich
magma in volcanoes (e.g. see Huppert \& Hallworth 2007 and references
herein) to plug-cementing oilfields (e.g. Frigaard \& Scherzer 1998,
Moyers-Gonzalez \& Frigaard 2004). There is also associated work on exchange problems involving miscible fluids, tilted tubes or channels, and unsteady solutions (see the recent articles by Seon et al. 2007, Znaien et al. 2009 and Taghavi et al. 2009 for references).

Resolving the flow degeneracy of the steady state in favour of one
realised solution involves knowledge of the initial conditions of the
exchange flow, the pressure boundary conditions set-up across the tube
and the inherent instability mechanisms present.  Pragmatically, the
initial conditions are never known that well (e.g. barriers are slid
open or plugs removed in the laboratory), the pressure gradient which
gets set up difficult to measure and assessing relative stability
requires every possible flow state to be identified first. It is
therefore tempting to jump to an ad-hoc selection principle especially
as a particularly obvious one suggests itself here: {\em the flow
  selects the solution which has the largest individual volumetric
  flux}.  A selection principle based upon maximum
flux has some history in the undirectional two-phase flow problem
motivated by its formal connection to the single fluid problem
(Maclean 1973, Everage 1973, Joseph, Nguyen \& Beavers 1984). Here,
the governing Stokes equations are the Euler-Lagrange equations for
maximising the flux for velocity fields which satisfy the global power
balance that the rate at which energy is viscously dissipated equals
the power supplied by the applied pressure gradient (per unit length
of the tube).  Specifically, if $G$ is the constant applied pressure
gradient, $\Omega$ the cross-section of the tube and $u$ the speed
along the tube, then
\beq
\mu \nabla^2 u=G \qquad
\Leftrightarrow \qquad \delta
\int_{\Omega} \, u+\Lambda (\mu |\nabla u|^2 +Gu)\,dA =0
\label{variational}
\eeq
where $\delta$ indicates the Frech\'{e}t (variational) derivative,
$\int\, -Gu\, dA$ is the rate of working by the pressure gradient per
unit length of tube and the Lagrange multiplier $\Lambda$ imposing the
power balance constraint takes the value $1/G$.  The stationary point
defined by the variational solution is clearly one of maximum flux
because the only quadratic term in the integrand is negative definite
($u$ is oppositely signed to $G$ so $\Lambda<0$)%
\footnote{Due to the relative simplicity of Stokes equations, there
  are many other variational formulations such as {\em maximising} the
  dissipation subject to the global power balance, {\em minimising}
  the dissipation subject to fixed flux and the complementary problem
  of {\em maximising} the flux subject to fixed dissipation.}.
The fact that this variational formulation can be extended to two
fluids {\em provided} the interface between them is known (Maclean
1973, Everage 1973) supplied the impetus to invoke the principle of
maximal flux more generally. It appears to be mostly successful - in
the words of Joseph, Nguyen and Beavers (1984) ``our experiments show
that something like this is going on''- predicting that the more
viscous fluid will be encircled by the less viscous fluid which then
acts as a lubricant against the tube walls (see also Charles \&
Redberger 1962, Yu \& Sparrow 1967, Hasson, Mann \& Nir 1970, Southern
\& Ballman 1973, Everage 1973, Lee \& White 1974, Williams 1975,
Minagawa \& White 1975).  Joseph, Renardy \& Renardy (1984), however,
add some qualifications: this state can become unstable if the more
viscous core gets too small.

Given this history, the purpose of this paper is to explore the
consequences of this `maximum flux principle' in predicting the form
of the exchange flow realised in a vertical cylindrical tube. Formally
solving the variational problem with the interface (or interfaces) as
an unknown is a formidable challenge not attempted here. Rather, a
survey is conducted over a physically-motivated subspace of all
mathematically-possible steady, coaxial solutions. This subspace is
defined by two (mild) assumptions: a) the fluids occupy one (possibly
multi-connected) domain so that there is only one interface $\Gamma$,
and b) that this interface is a circular arc or a full circle.  The
motivation for the former assumption is stability - multiple small
fluid domains would presumably aggregate - and the presence of some
surface tension between the two fluids conveniently motivates the
latter. The axially-constant, lateral pressure difference required to balance
interfacial tension, however, will be ignored in what follows as it
has no consequence for the calculations.

\section{Formulation}

Consider two immiscible fluids with densities $\rho_1$ and $\rho_2$
and viscosities $\mu_1$ and $\mu_2$ which are flowing in a vertical
circular tube of radius $a$ across which there is a pressure gradient
$G$ and $g$ is the acceleration due to gravity. Assuming that fluid
1(2) occupies an area $A^*_1$($A^*_2$), the Navier-Stokes equations
for steady exchange flow of the two fluids either directed up or down
the tube (so the problem is just in the cross-sectional plane) are
\beq
G=\mu_1 \nabla^2 u^*_1 -\rho_1 g \quad {\rm in} \quad A^*_1, \qquad
G=\mu_2 \nabla^2 u^*_2 -\rho_2 g \quad {\rm in} \quad A^*_2
\eeq
with non-slip boundary conditions at the tube wall and continuity of
velocity and stress at the interface $\Gamma^*$ between the two fluids,
that is
\beq
u^*_1=u^*_2 \quad \& \quad
\mu_1 \frac{\partial u^*_1}{\partial n}= \mu_2
\frac{\partial u^*_2}{\partial n} \quad {\rm on \quad \Gamma^*}
\eeq
(where $\partial/\partial n$ is the normal derivative to $\Gamma^*$).
There is a further constraint that the net volume flux through the
tube is zero so
\beq
Q^*:=-\int \, u^*_1 \, dA^*_1 = \int \, u^*_2 \, dA^*_2.
\label{balance}
\eeq
Without loss of generality, we assume $\rho_1 > \rho_2$ so that $Q^*$ is
positive (the less dense fluid rises). This does not prejudice the choice
of viscosities later because of the symmetry
$(\rho_1,\rho_2,g) \rightarrow (\rho_2,\rho_1,-g)$: the direction `up' is irrelevant with only the density difference being important.

The system is non-dimensionalised (*'s removed) using the
tube radius $a$, the differential hydrostatic pressure gradient
$\Delta \rho g$ (where $\Delta \rho:= \rho_1-\rho_2$) and $\mu_1$ so
that after defining $\lambda$ by
\beq
G=-\half(\rho_1+\rho_2)g+\half \Delta \rho g \lambda
\eeq
then
\begin{eqnarray}
       \nabla^2 u_1 &=& \lambda+1 \quad {\rm in} \quad A_1, \label{prob1}\\
\beta \nabla^2 u_2 &=& \lambda-1 \quad {\rm in} \quad A_2, \label{prob2}
\end{eqnarray}
\beq
u_1=u_2 \quad  \& \quad
\frac{\partial u_1}{\partial n}= \beta
\frac{\partial u_2}{\partial n} \quad {\rm on \quad \Gamma}.  \label{bcs}
\eeq
where
\beq
\beta:=\frac{\mu_2}{\mu_1}.
\eeq
Henceforth $u_1$ and $u_2$
are in units of $\half \Delta \rho g a^2/\mu_1$ and the one-fluid
volume flux
\beq
Q:=-\int \, u_1 \, dA_1 = \int \, u_2 \, dA_2
\eeq
is in units of $\half \Delta \rho g a^4/\mu_1$ with $A_1 \cup A_2$
being the unit disk.

%
%
\begin{figure}
\centering
\resizebox{0.9\textwidth}{!}{\includegraphics{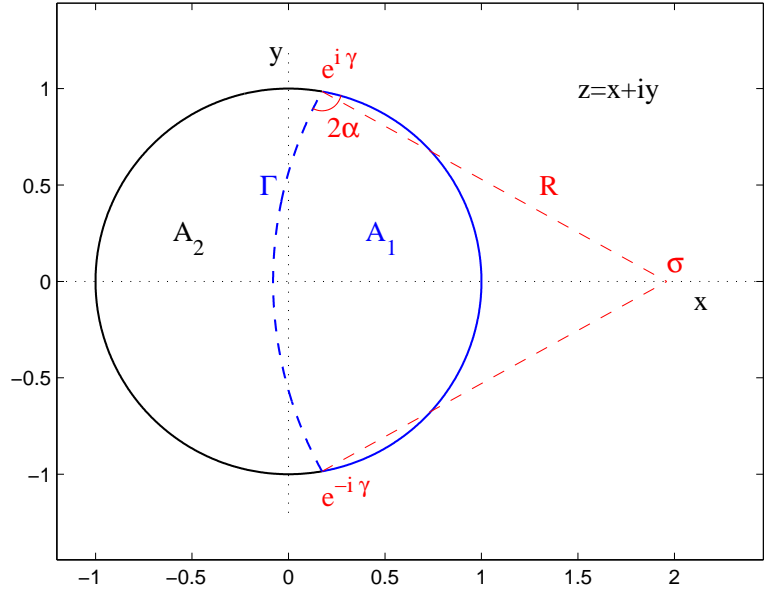}}
\caption{The side-by-side solution configuration specified by two
  parameters: $\gamma$ and $\alpha$.}
\label{zdiagram}
\end{figure}

%
%
\begin{figure}
\centering
\resizebox{0.9\textwidth}{!}{\includegraphics{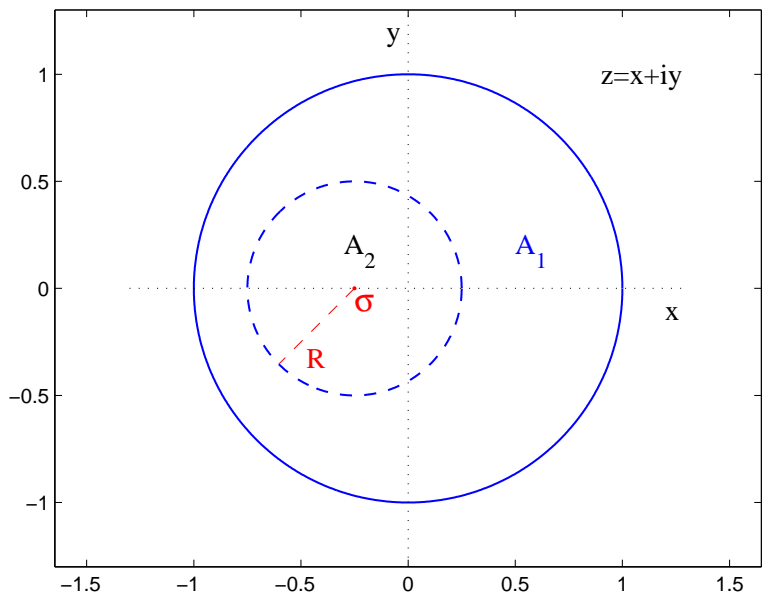}}
\caption{The eccentric core-annular configuration specified by two
  parameters: $\sigma$ and $R$.}
\label{zdiagram1}
\end{figure}

Two specific choices are now made for $\Gamma$. The first is a
circular arc of general curvature and position which intersects the
tube wall so that the two fluids are next to each other - the {\em
  side-by-side} solution: see figure \ref{zdiagram}. The second is a
full circle completely contained within, but not concentric with, the
tube so that one fluid encapsulates the other - the {\em eccentric}
core-annular solution: see figure \ref{zdiagram1}. The limiting case
of a {\em concentric} core-annular solution needs to be treated
separately but is easily solved analytically.\\
 
\subsection{Side-by-side solutions} 

The geometry of the side-by-side solution is shown in figure
\ref{zdiagram} to be defined by two parameters: $\gamma$, the (upper)
intercept latitude of $\Gamma$ with the tube wall, and $2\alpha$, the
angle between $\Gamma$ and tube wall.  For given viscosity ratio
$\beta$ and pressure gradient $\lambda$, one of these (nominally
$\alpha$) is determined by the flux balance leaving a 1-dimensional
family of side-by-side flows with corresponding fluxes
$Q=Q_s(\beta,\lambda;\gamma)$ possible (see appendix A for the
calculation details). There is a symmetry
\beq
Q(\beta,\lambda;\gamma,\alpha)=\frac{1}{\beta}Q(\frac{1}{\beta},-\lambda;\pi-\gamma,\frac{\pi}{2}-\alpha)
\eeq
which means that only $\beta \geq 1$ need be considered providing the
full ranges of $\gamma$ and $\alpha$ are studied.  Henceforth fluid 2
will always be the more viscous fluid so that the
non-dimensionalisation has been done using the smaller dynamic
viscosity $\mu_1$.\\

\subsection{Eccentric solutions}

The eccentric core-annular solution has one
fluid domain as a totally-contained circular disk (cylinder) not
touching the tube wall. The radius $R<1$ and centre $(\sigma,0)$ of
$\Gamma$ define the geometry uniquely up to obvious rotations and
reflections. To match smoothly onto the choices made in the
side-by-side solution, $\sigma$ is chosen to be +ve(-ve) for $A_1$ in
$A_2$ ($A_2$ in $A_1$). As before, for given viscosity ratio $\beta$ and
pressure gradient $\lambda$, one of these two geometrical parameters
is determined by the flux balance.  This is done by searching over $R$
for given
\beq
d:=\biggl\{ 
\begin{array}{rl}
 1+\sigma-R &  \qquad  A_2 \quad {\rm in} \quad A_1 \qquad \sigma <0\\
-1+\sigma+R & \qquad   A_1 \quad {\rm in} \quad A_2 \qquad \sigma >0
\end{array}
\biggr.
\eeq
which either represents the positive displacement from $(-1,0)$ to
$(\sigma-R,0)$, the leftmost point of $\Gamma$ for the case of $A_2$
in $A_1$ ($\sigma<0$) , or the negative displacement of
$(\sigma+R,0)$, the rightmost point of $\Gamma$, from $(1,0)$ for the
case of $A_1$ in $A_2$ ($\sigma>0$). This choice is made for two
reasons. Firstly, $d$ is a convenient way of extending the
side-by-side solutions continuously beyond their pinch-off points into
the corresponding eccentric solutions: $\gamma \rightarrow 0$
corresponds to $A_2$ encapsulating $A_1$ and $d$ decreasing across
zero whereas $\gamma \rightarrow \pi$ corresponds to $A_1$
encapsulating $A_2$ and $d$ increasing across zero (see figure
3). Secondly, only one flux-balanced solution was ever found for a
given $d$ whereas some $\sigma$ can have two flux-balanced
solutions. The result is that two 1-dimensional families of eccentric
core-annular flows with corresponding fluxes $Q_e(\beta,\lambda;d)$
(more viscous core) and $\hat{Q}_e(\beta,\lambda,d)$ (less viscous
core) are possible (see appendix B for the calculation details). It's
worth re-emphasizing here that $\beta \geq 1$ so all the flux values
quoted are in units of $1/\mu_1$ where $\mu_1$ is the smaller dynamic
viscosity.

\subsection{Concentric solutions} 

When $\Gamma$ is a circle concentric with
the tube wall there is a simple solution to the problem
(\ref{prob1})-(\ref{bcs}) discussed recently by Huppert \& Hallworth
(2007):
\begin{eqnarray}
u_1 &=& \frac{\lambda+1}{4}(r^2-1)-R^2 \log r,  \hspace{3cm}  R \leq r\leq 1\\
u_2 &=& \frac{\lambda-1}{4 \beta}(r^2-R^2)-R^2\log R-\frac{\lambda+1}{4}(1-R^2).
\qquad r \leq R
\end{eqnarray}
The associated fluxes are
\begin{eqnarray}
Q_1 &=& \frac{\pi}{8}
\biggl[ (\lambda+1)(2R^2-R^4-1)+4R^2(1-R^2)+8R^4 \log R \biggr], \label{Q1}\\
Q_2 &=& \frac{\pi}{8 \beta}
\biggl[  (1-\lambda)R^4
-2 \beta(1+\lambda)R^2(1-R^2)-8\beta R^4 \log R
\biggr]. \label{Q2}
\end{eqnarray}
Since this is a special case of an eccentric core-annular solution
with $\sigma=0$, there is unique $0<R<1$ for a flux-balanced
solution which is
\beq R_c=\sqrt{\frac{2\beta-\sqrt{4 \beta^2-
      \beta(1+\lambda)[\beta(3-\lambda)+(\lambda-1)]}}
  {[\beta(3-\lambda)+(\lambda-1)]}}.
\eeq
so that the flux (for fluid 2 in the core) is $Q_c(\beta, \lambda)$. As $\beta \rightarrow
\infty$, 
\beq
R_c \rightarrow
\sqrt{(1+\lambda)/(3-\lambda)}, \qquad \lambda \rightarrow 0.1746 \qquad {\rm and} \quad Q_c \rightarrow 0.01831 
\eeq
from above. The opposite scenario of the less
viscous fluid (fluid 1) in the core has
$Q:=\hat{Q}_c\sim O(\beta)$ ($\beta \rightarrow 1/\beta$ in
expressions (\ref{Q1}) and (\ref{Q2}) and multiply $Q$ by $1/\beta$ to convert the flux units to those using the smaller dynamic viscosity).

\subsection{Strategy}

The strategy now is to calculate $max_{\lambda}Q$ as a function of
$\beta$ over all possible geometries smoothly ranging from the
concentric solution with {\em less} viscous fluid in the core through
to the concentric solution with the {\em more} viscous fluid in the
core.  Figure 3 illustrates the spectrum of possibilities and a
glimpse of how the flux varies at one $\beta$ value. Before detailing
the results further, the reader may be amused by an admission. At
onset, this author (naively?) expected the calculation of maximum flux
to be a simple competition between a local maximum achieved by the
side-by-side solution and the flux $Q_c$ associated with the
concentric core-annular flow influenced by the known
behaviour of unidirectional 2-fluid flow. The side-by-side solution,
however, quickly loses its interior maximum ($0<\gamma <\pi$) as
$\beta$ increases in favour of an end-point maximum at
$\gamma=\pi$. The fact that this end-point maximum {\it exceeds} the
concentric solution flux $Q_c$ unequivocally indicated the importance
of the intermediate eccentric core-annular flux $Q_e$.

%
                                                                                
\begin{figure}                                                                  
\setlength{\unitlength}{1cm}                                                    
\begin{picture}(14,14)                                                          
\put(-1,0){\includegraphics[width=16cm]{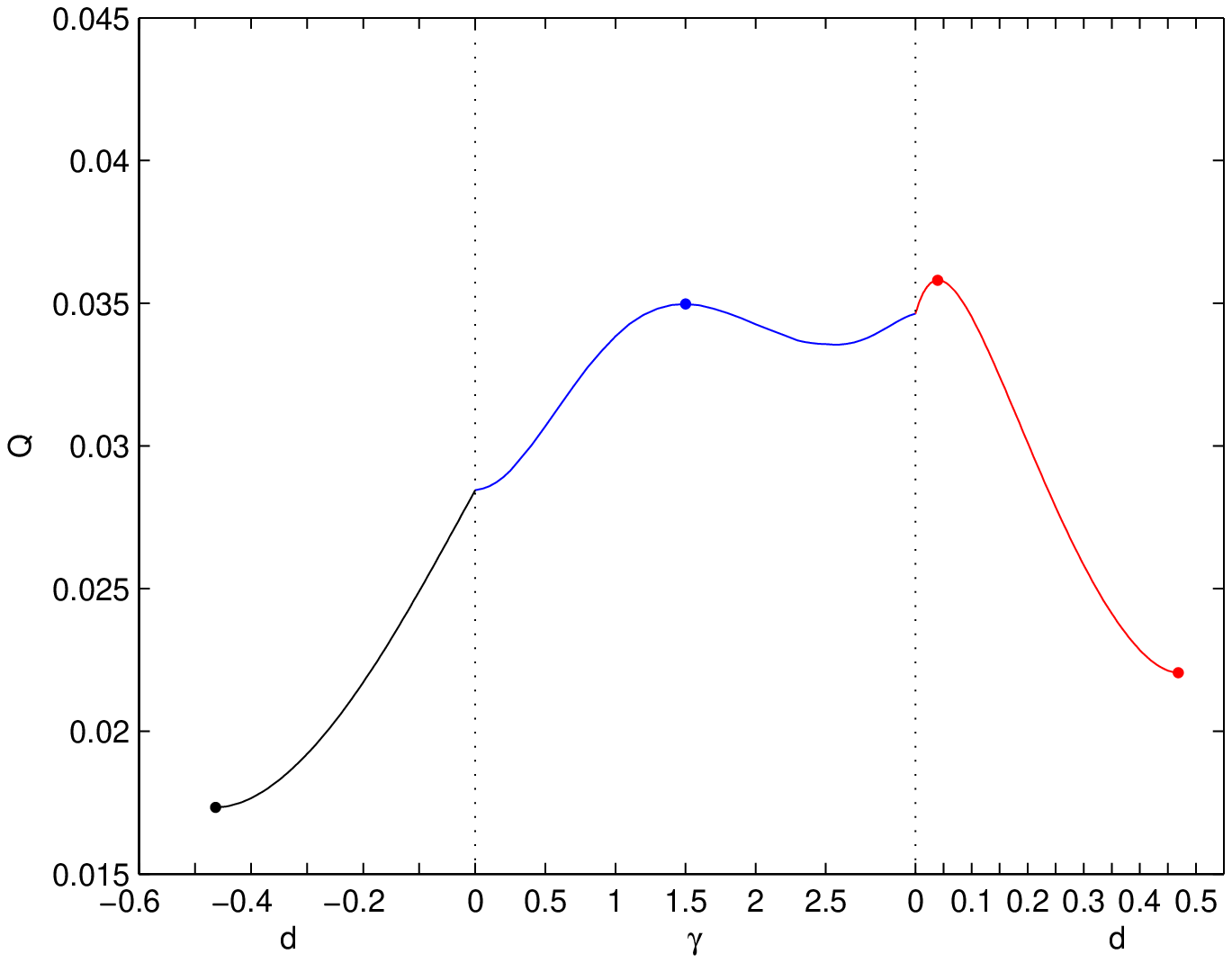}}
\put(1.25,9.5){\includegraphics[width=1.75cm]{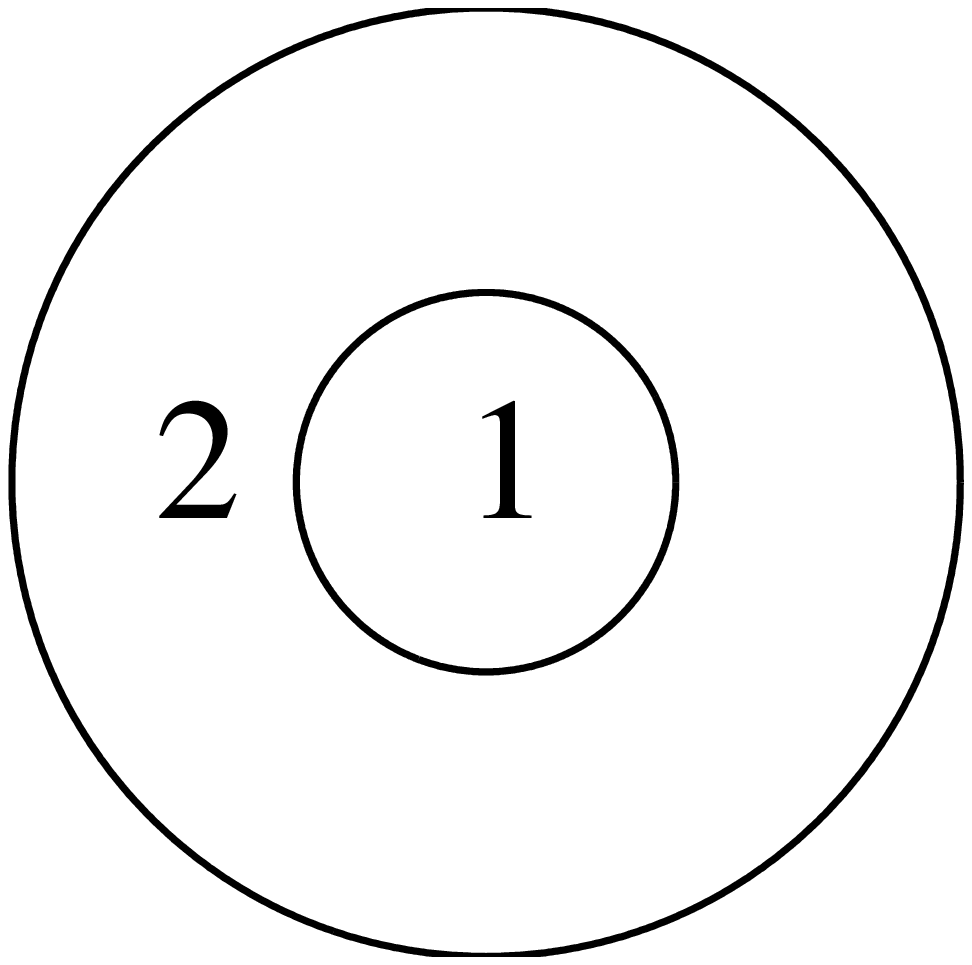}}
\put(3,9.5){\includegraphics[width=1.75cm]{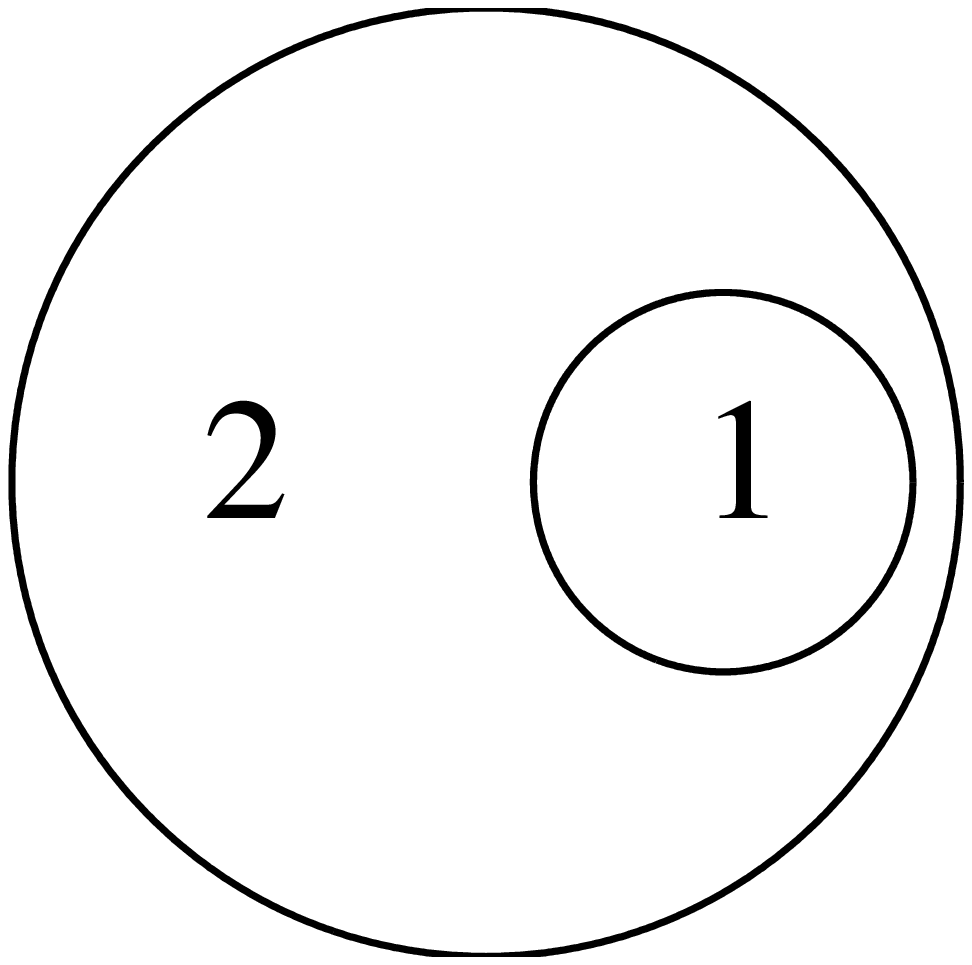}}
\put(5,9.5){\includegraphics[width=1.75cm]{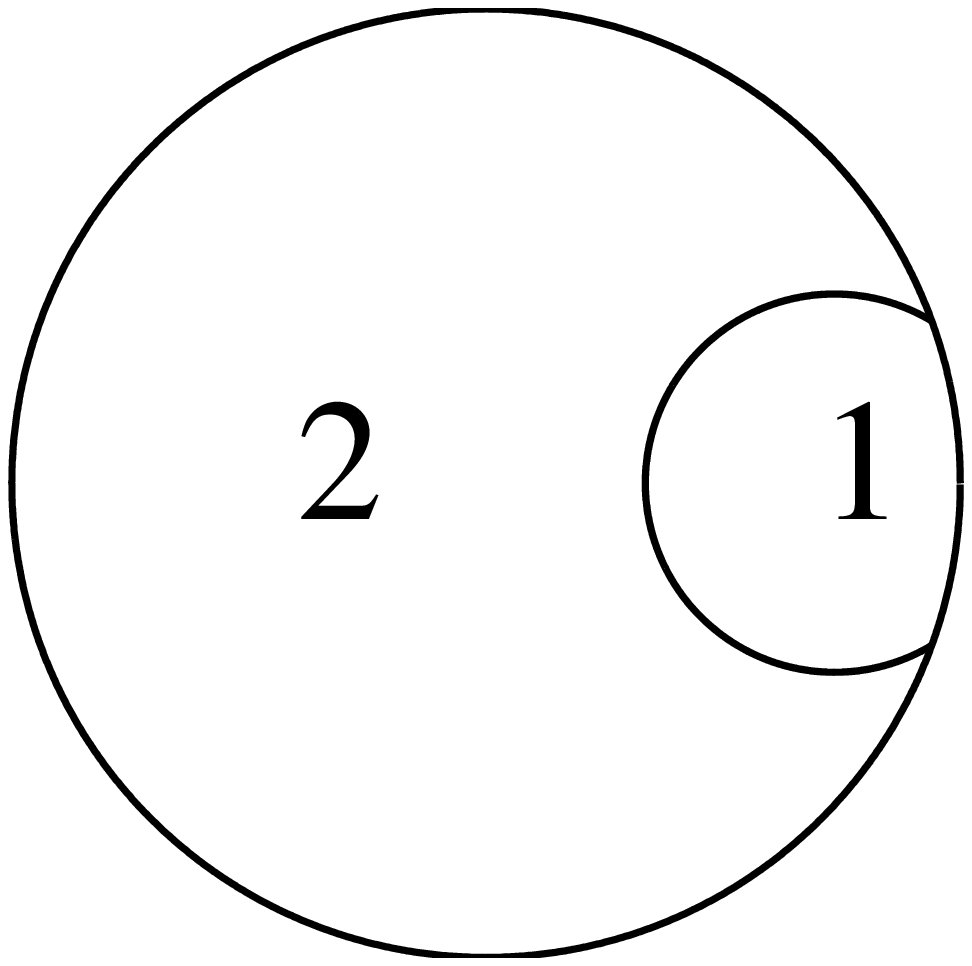}}
\put(6.5,9.5){\includegraphics[width=1.75cm]{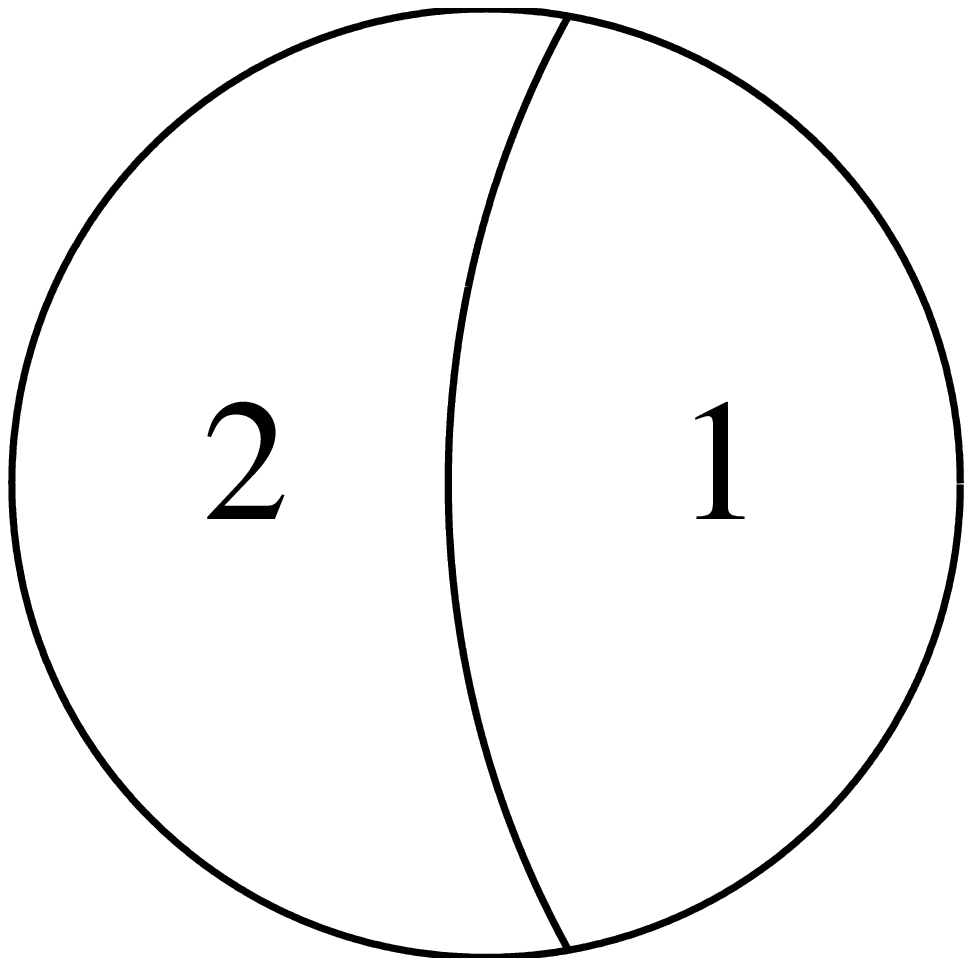}}
\put(8,9.5){\includegraphics[width=1.75cm]{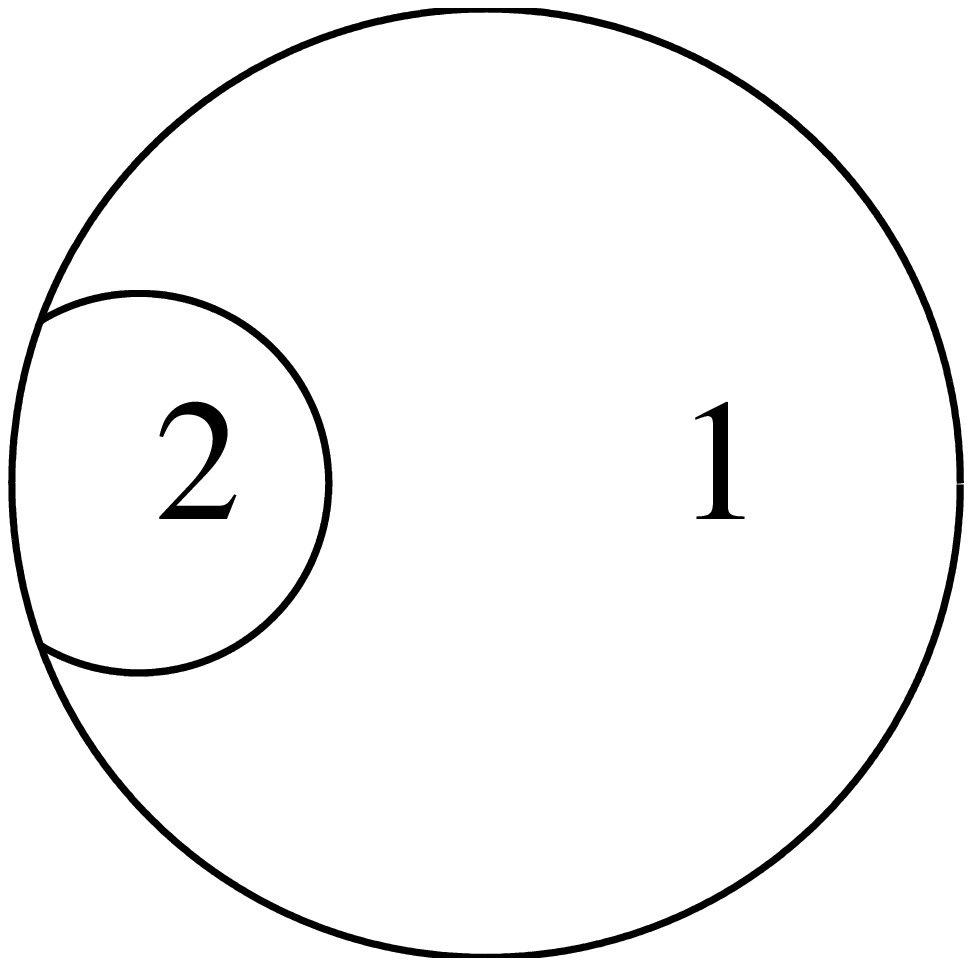}}
\put(10,9.5){\includegraphics[width=1.75cm]{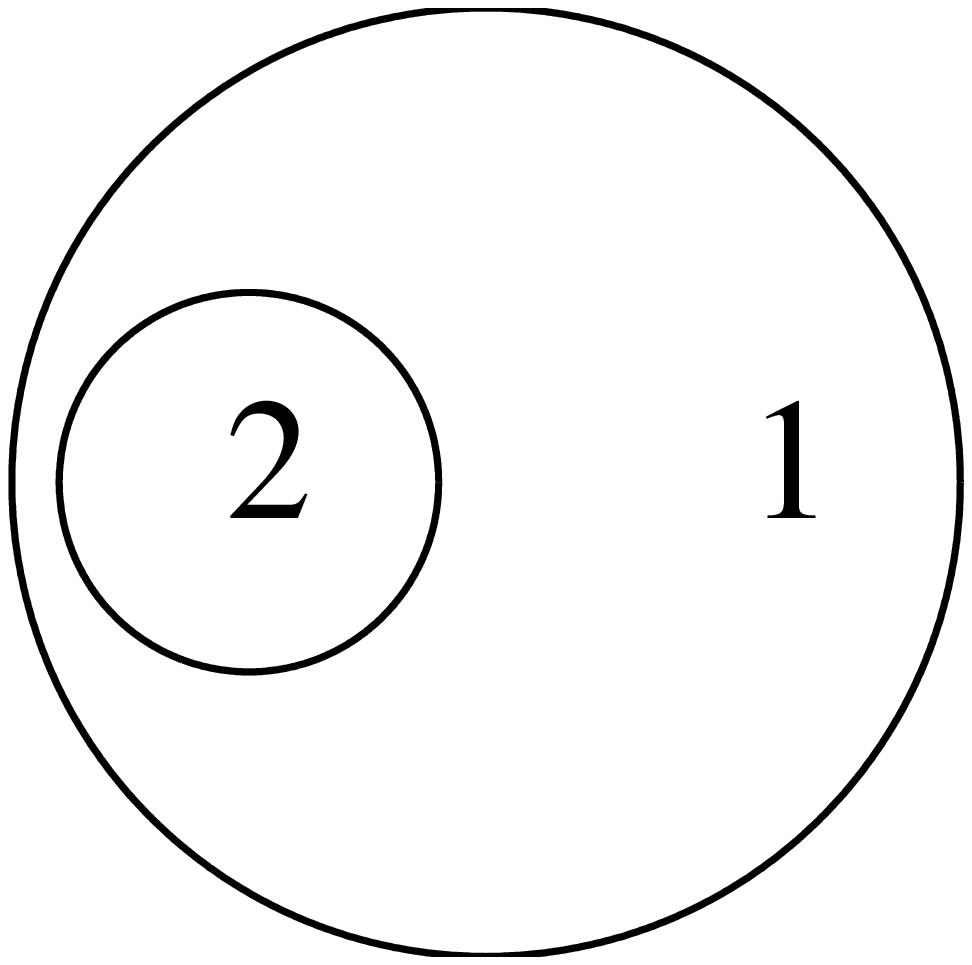}}
\put(11.5,9.5){\includegraphics[width=1.75cm]{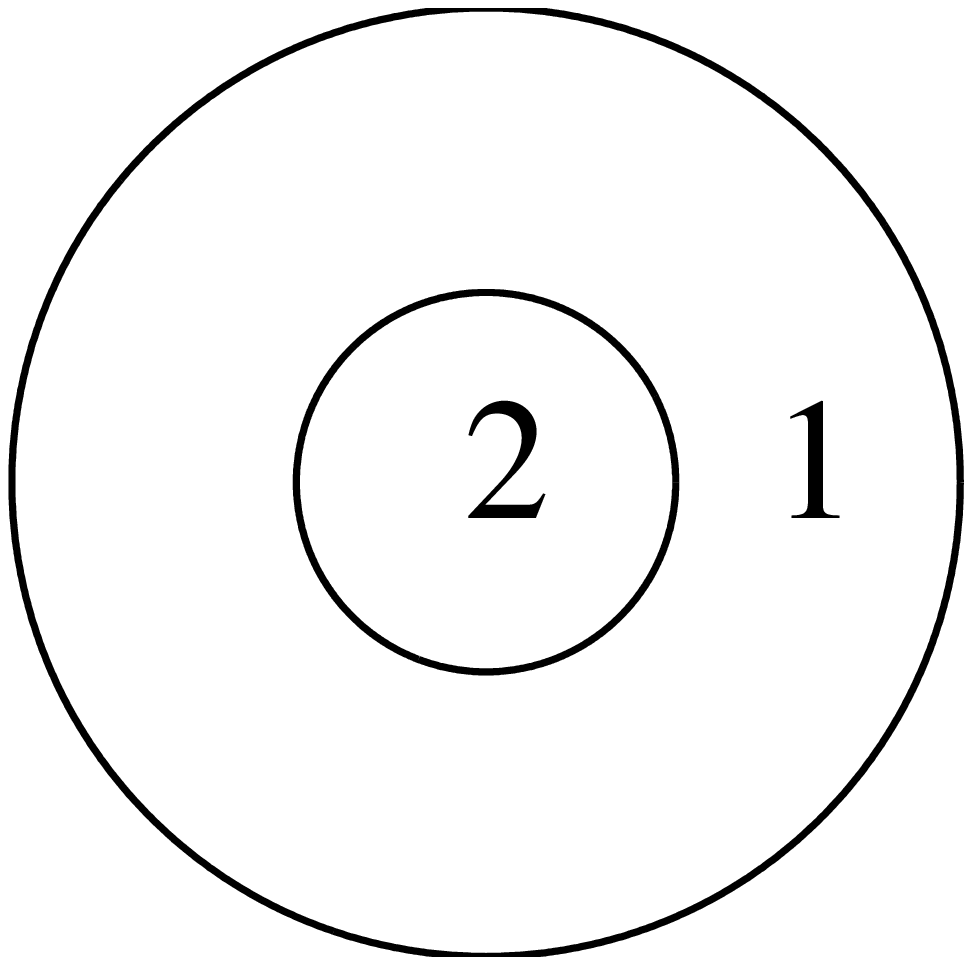}}
\end{picture}                                                                   
\caption{The flux ${\rm max}_{\lambda}Q$ plotted across the various
  flow configurations ($d<0$ indicates less viscous core and $d>0$
  more viscous core) for $\beta=5$. The leftmost point is $\hat{Q}_c$,
  beyond this, the region $d<0$ is the domain for $\hat{Q}_e$, the
  region $\gamma \in [0,\pi]$ is the domain for $Q_s$, $d>0$ the
  domain for $Q_e$ and the rightmost point is $Q_c$. The interior
  local maxima are highlighted with dots. The curve is only $C^0$
  because the abscissa changes character at $\gamma=0$ and $\pi$ of
  course.}
  \label{intro}                                                                 
\end{figure}                                                                    

%
\begin{figure}
\centering
\resizebox{0.9\textwidth}{!}{\includegraphics{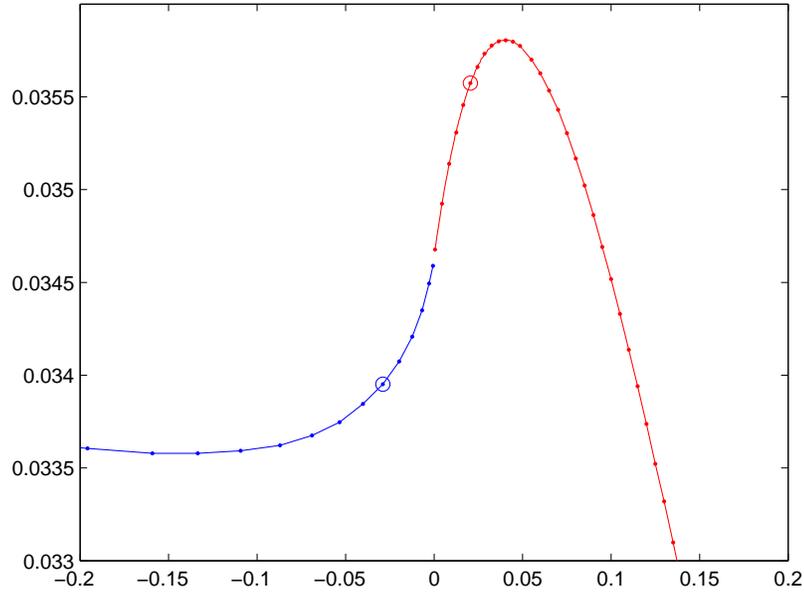}}
\caption{Plotting ${\rm max}_{\lambda} Q$ against $d$ at $\beta=5$
for side-by-side solutions as $d \rightarrow 0^{-}$ ($\gamma
\rightarrow \pi$) and eccentric solutions as $d \rightarrow 0^+$
demonstrates the smooth connection between the two
formulations. Velocity fields for the circled points are shown in
figure \ref{detail_images}. } 
\label{detail}
\end{figure}

\begin{figure}
\setlength{\unitlength}{1cm}
\begin{picture}(14,8)
\put(-1.5,0){\includegraphics[width=9cm]{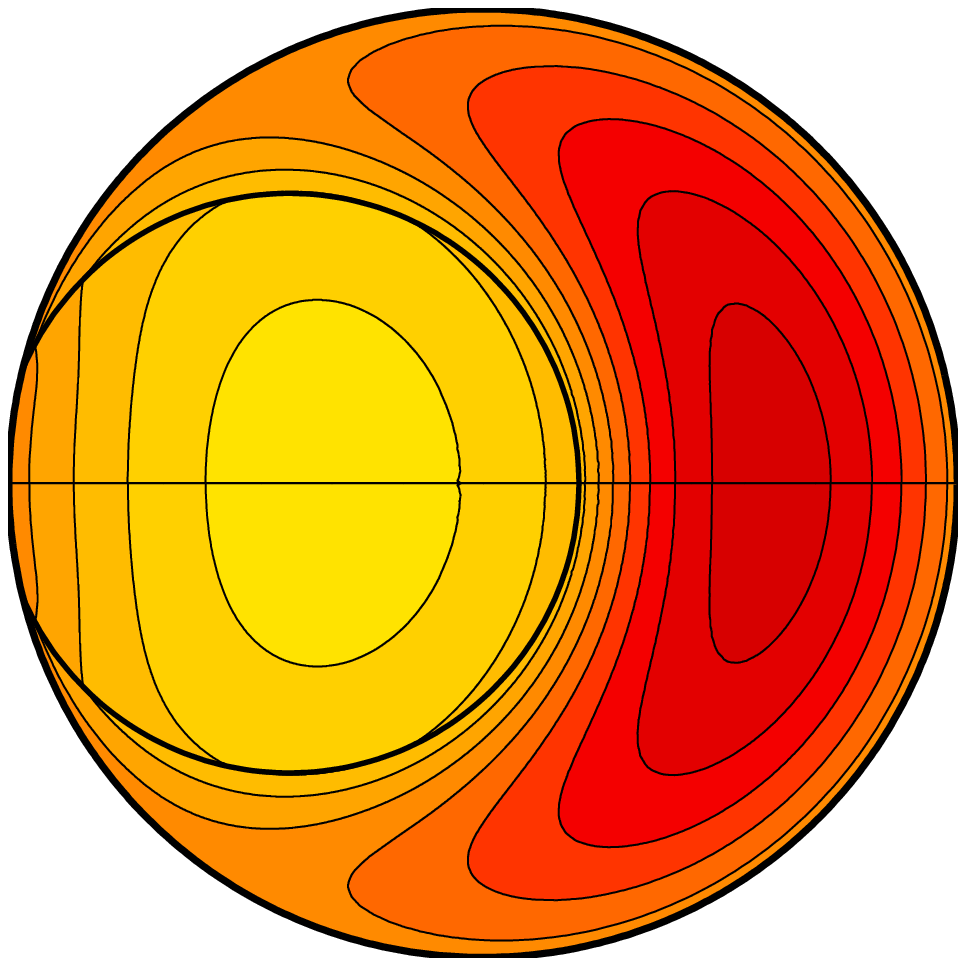}}
\put(6,0){\includegraphics[width=9cm]{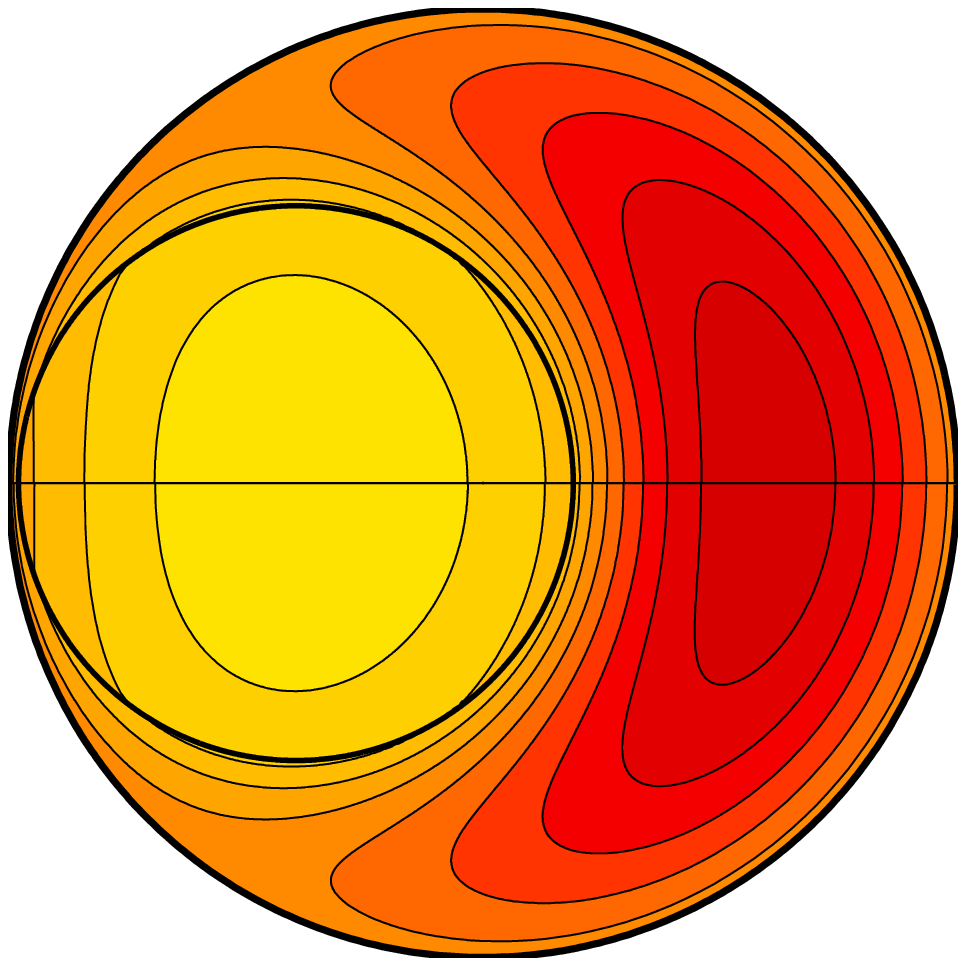}}
\end{picture}
\caption{The pinching-off side-by-side ($d=-0.0198$, $-0.051 \leq u
  \leq 0.041$) and near-touching eccentric solutions ($d=0.0205$,
  $-0.052 \leq u \leq 0.043$) for $\beta=5$ and the optimal
  $\lambda=-0.20$ corresponding to the circles in figure
  \ref{detail}. The contours range from -0.105 (dark/red) to 0.105
  (light/white) in steps of 0.01 here and throughout figures
  \ref{side} and \ref{eccentric} to aid
  comparison.} \label{detail_images}
\end{figure}

%
%
\begin{figure}
\centering \resizebox{\textwidth}{!}{\includegraphics{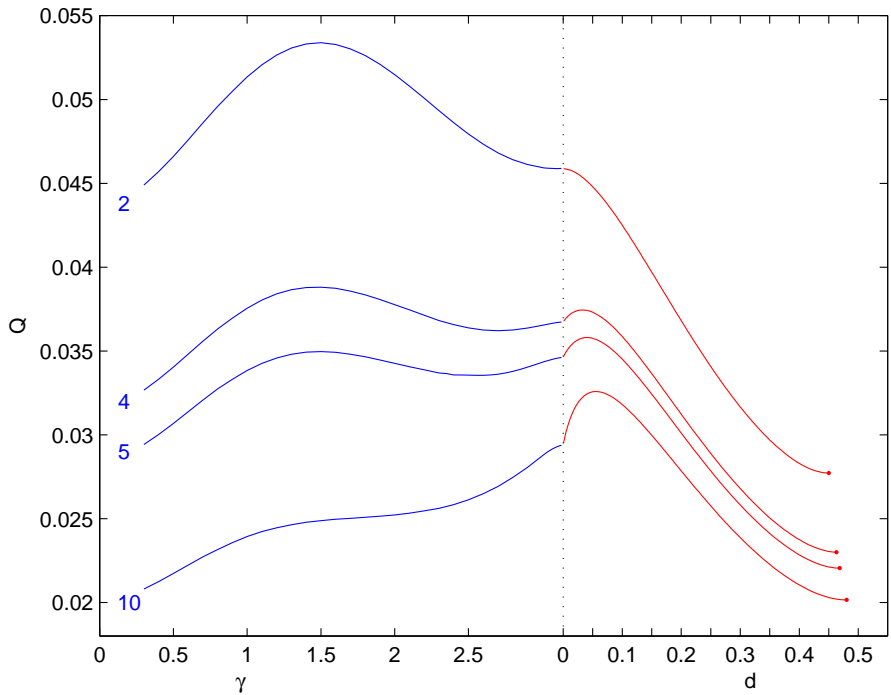}}
\caption{${\rm max}_{\lambda} Q$ as a function of $\gamma$ for the
side-by-side solutions and as a function of $d>0$ for the eccentric
solutions at $\beta=2,4,5$ and $10$. The single dot at the right end
of each curve corresponds to the concentric case $Q_c$. The global flux
maximum is a side-by-side solution for $\beta \leq 4.60$ and an
eccentric solution for $\beta \geq 4.60$.} \label{merge1}
\end{figure}

%
%
\begin{figure}
\setlength{\unitlength}{1cm}
\begin{picture}(14,14)
\put(-1.5,7){\includegraphics[width=9cm]{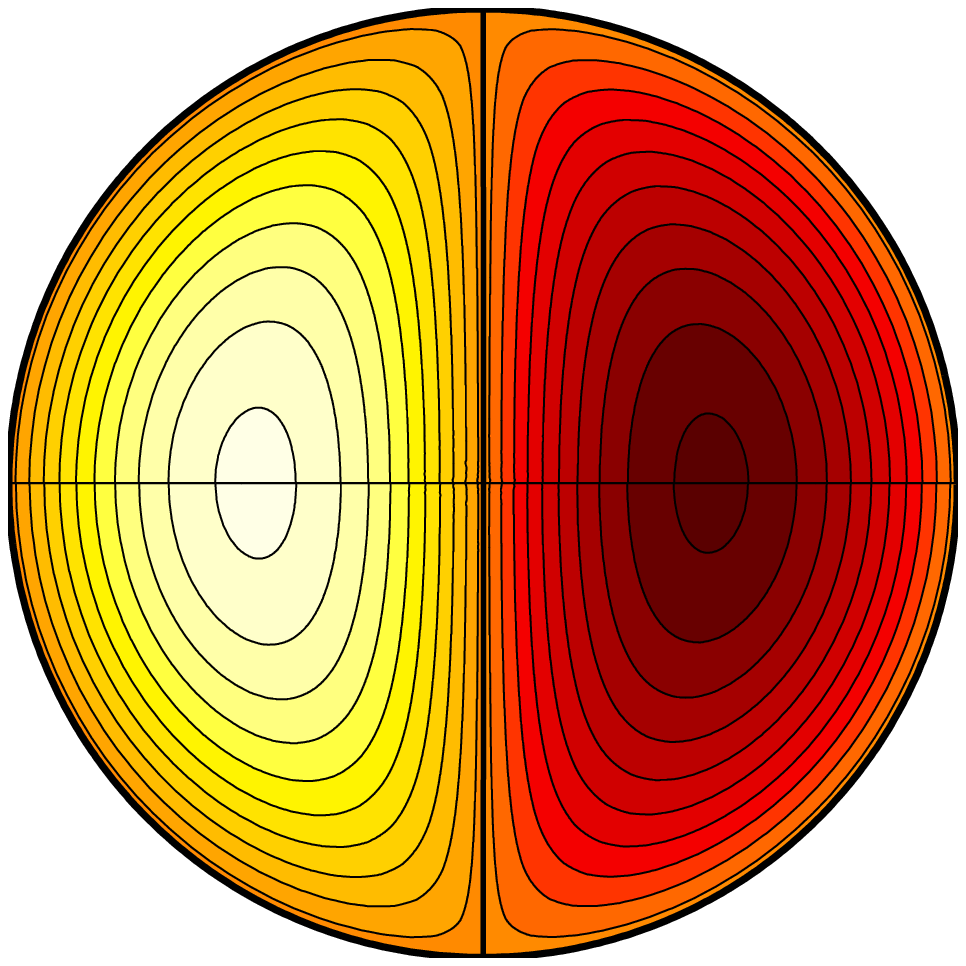}}
\put(6,7){\includegraphics[width=9cm]{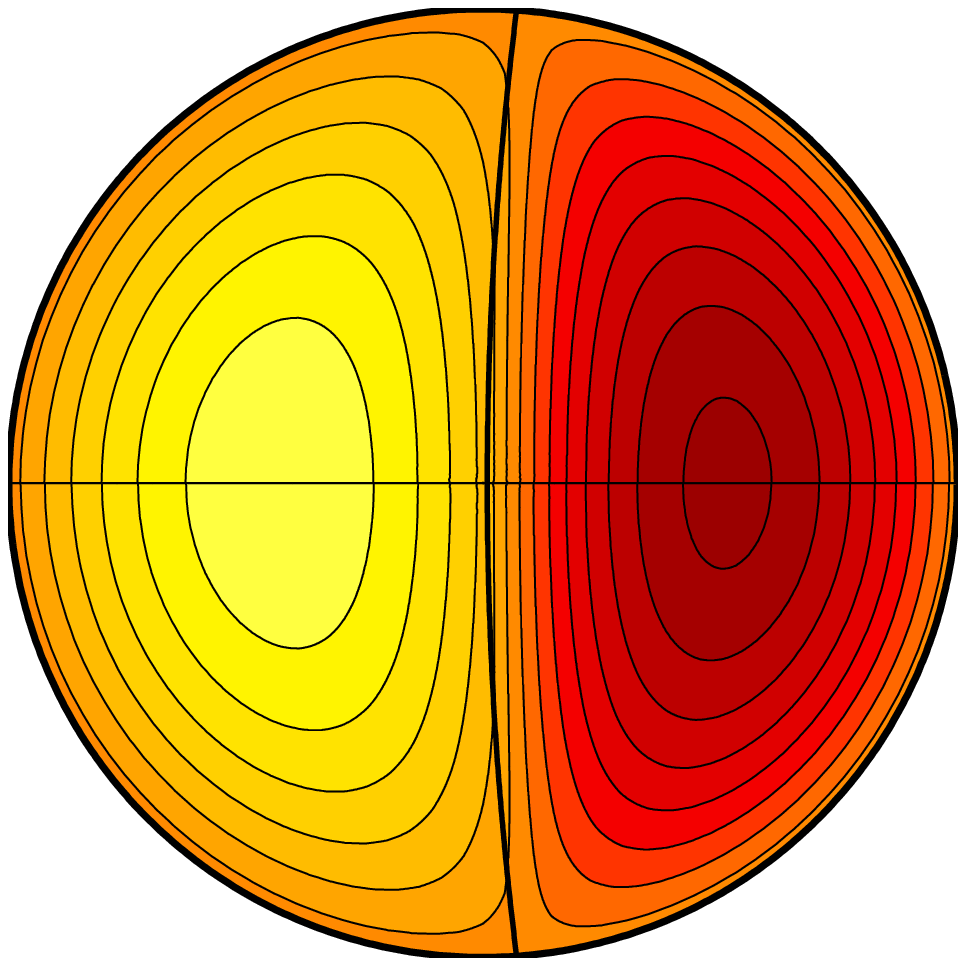}}
\put(-1.5,0){\includegraphics[width=9cm]{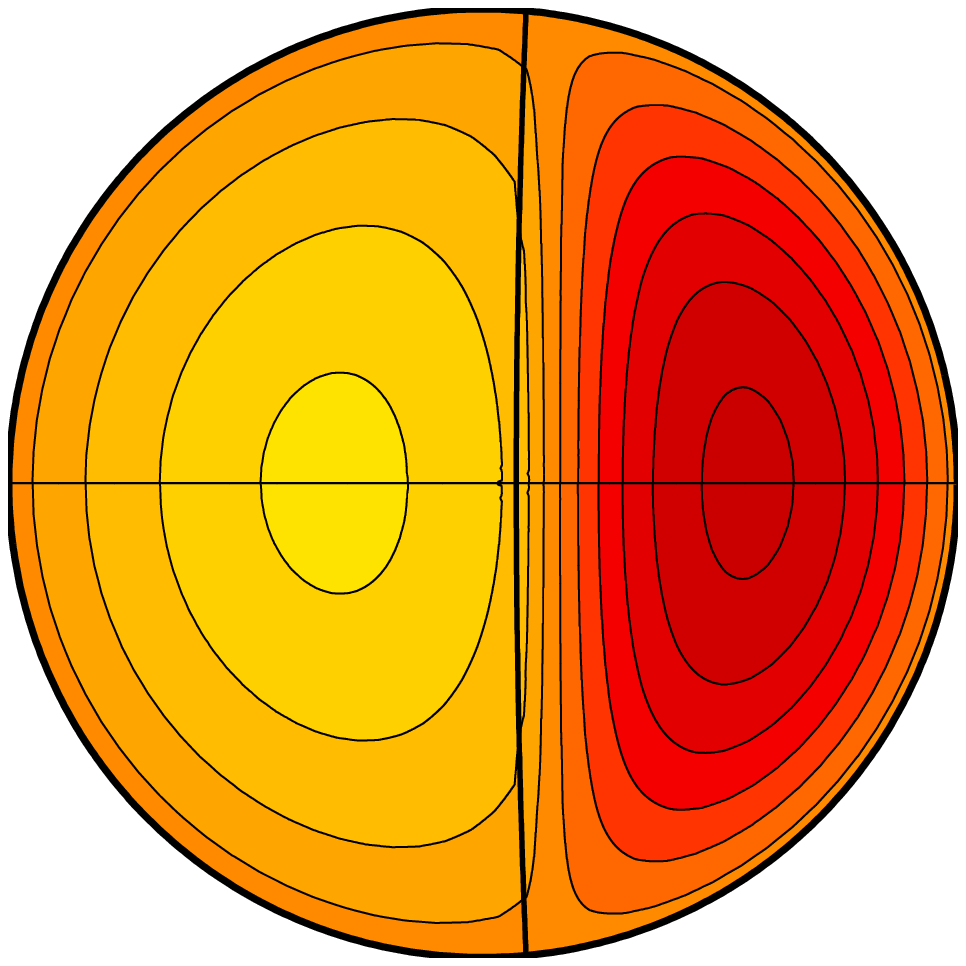}}
\put(6,0){\includegraphics[width=9cm]{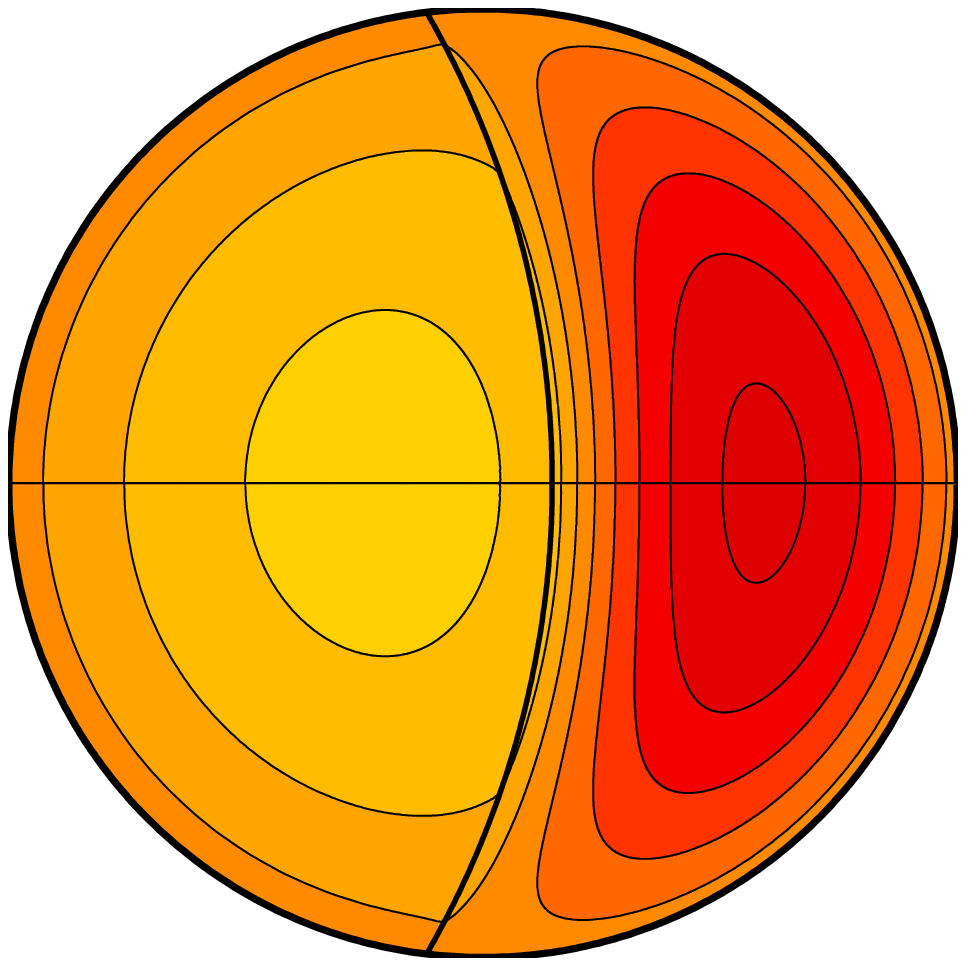}}
\end{picture}
\caption{Maximal flux side-by-side solutions for $\beta=1$ (top left,
 $-0.098 \leq u \leq -0.098$), $\beta=2$ (top right, $-0.078\leq
  u\leq 0.063$), $\beta=5$ (bottom left, $-0.058\leq u \leq 0.037$) 
and $\beta=8$
  (bottom right, $-0.047 \leq u \leq 0.029$). The contours range from -0.105(dark/red) to 0.105 (light/white) in steps of 0.01 here and throughout figures
  \ref{detail_images} and \ref{eccentric} to aid
  comparison.} \label{side}
\end{figure}
%
%
\begin{figure}
\setlength{\unitlength}{1cm}
\begin{picture}(14,14)
\put(-1.5,7){\includegraphics[width=9cm]{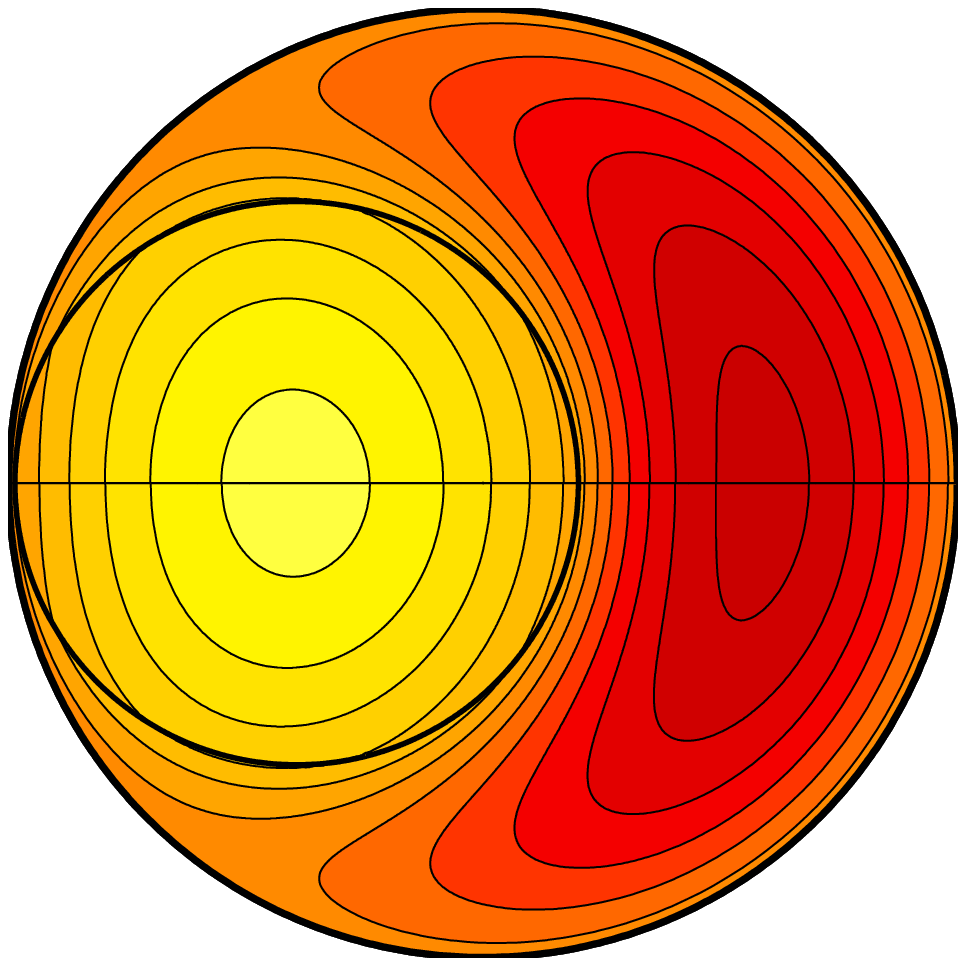}}
\put(6,7){\includegraphics[width=9cm]{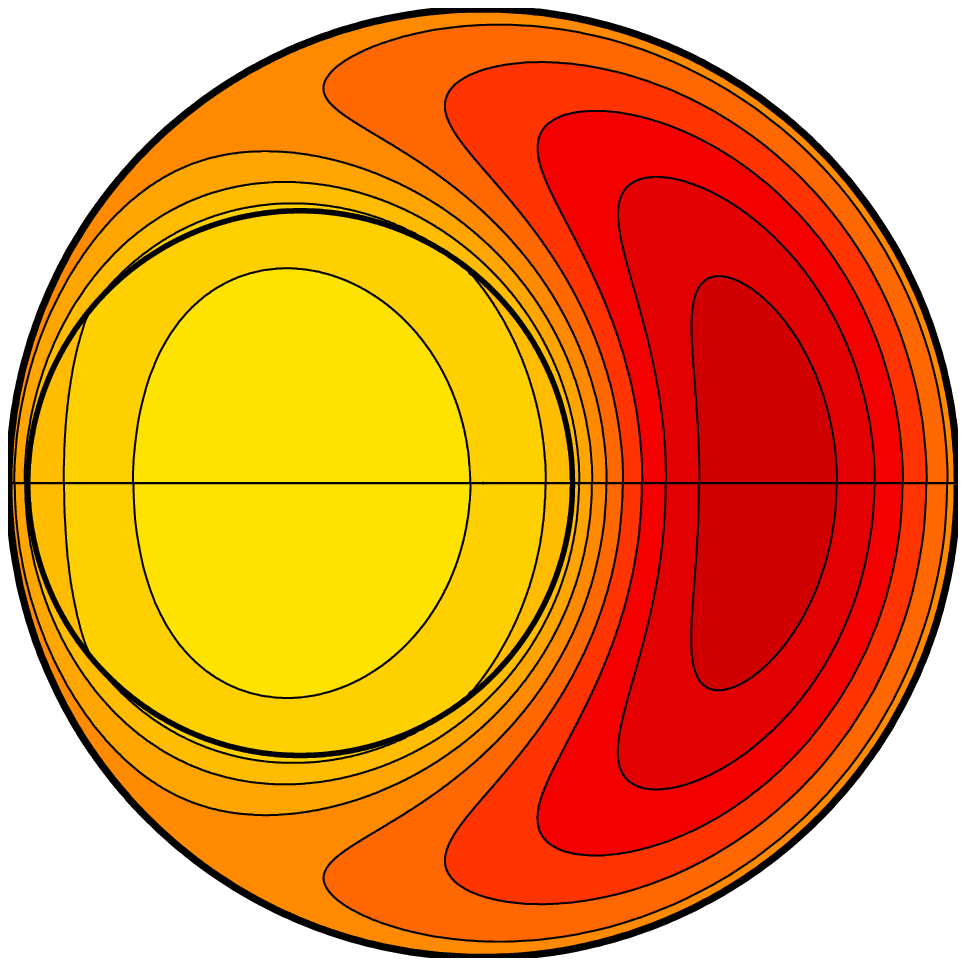}}
\put(-1.5,0){\includegraphics[width=9cm]{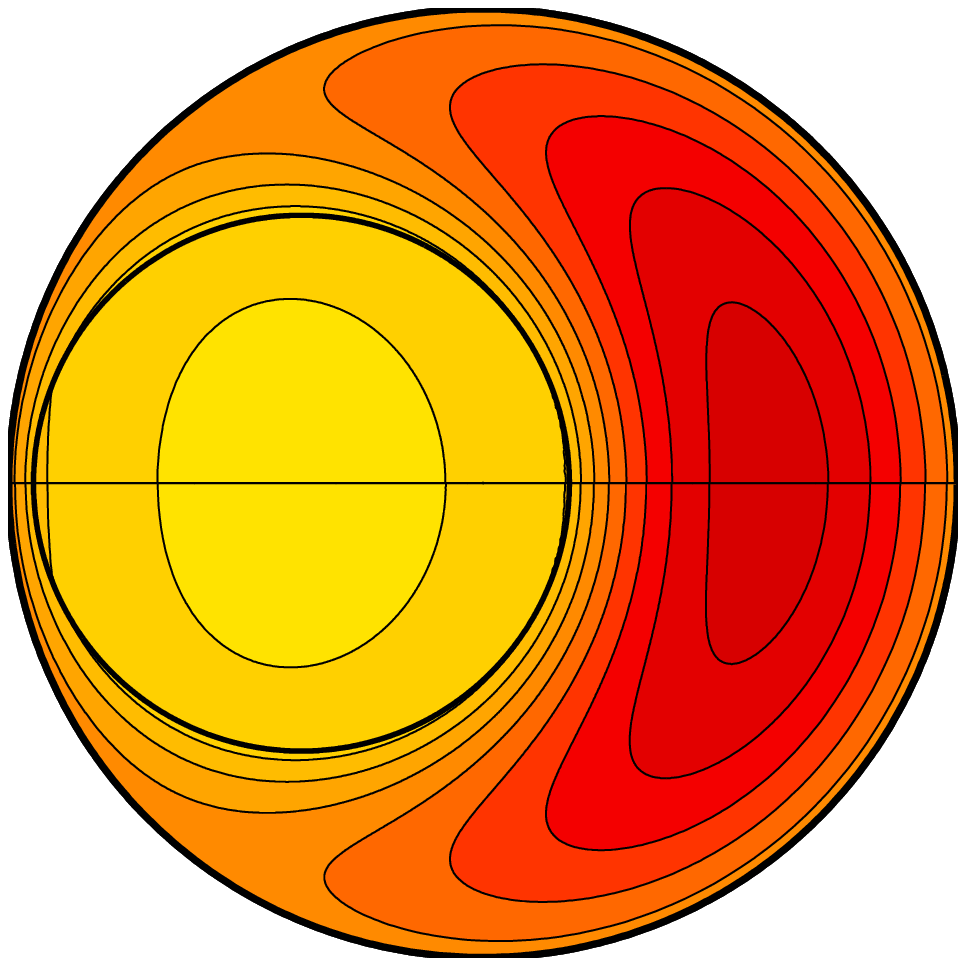}}
\put(6,0){\includegraphics[width=9cm]{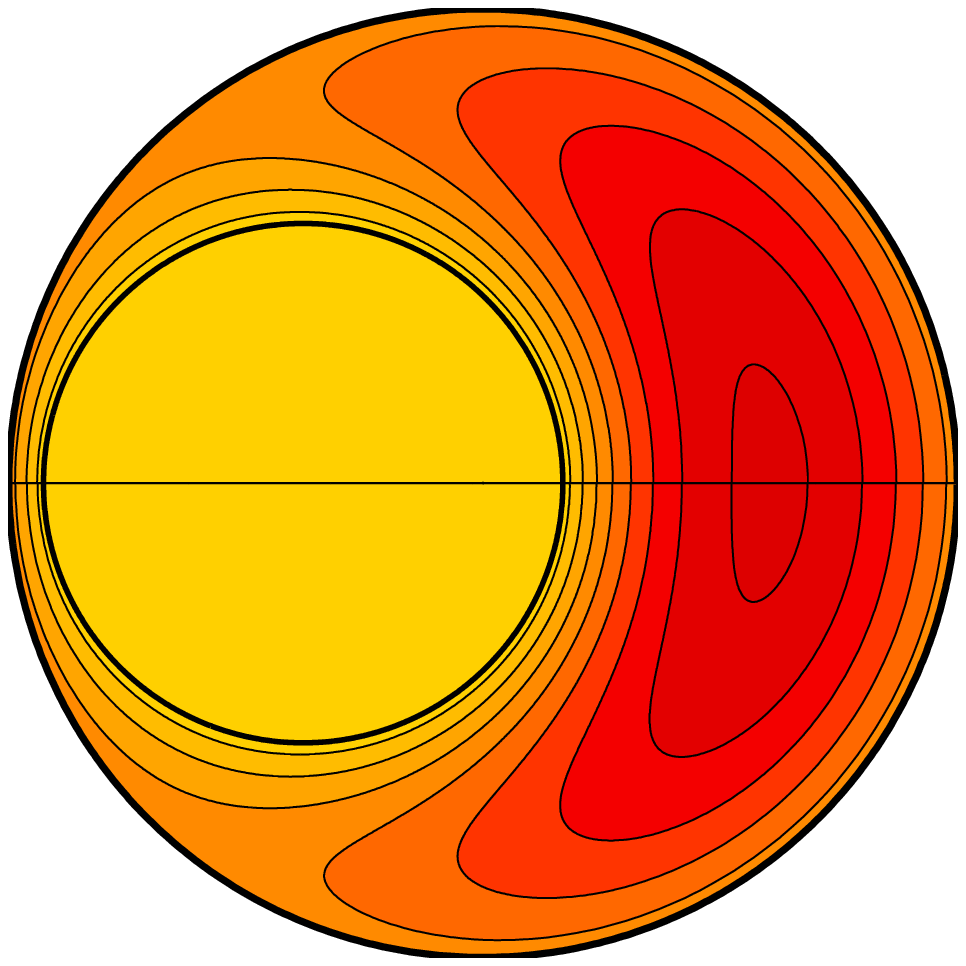}}
\end{picture}
\caption{Maximal flux eccentric core-annular solutions for $\beta=2.5$
  (top left, $-0.059 \leq u \leq 0.058$ ), $\beta=5$ (top right,
  $-0.053 \leq u \leq 0.044$), $\beta=8$ (bottom left, $-0.051 \leq u
  \leq 0.039$) and $\beta=10,000$ (bottom right, $-0.047 \leq u \leq
  0.031$). The contours range from -0.105(dark/red) to 0.105 (light/white) in steps of 0.01 here and throughout figures
  \ref{detail_images} and \ref{side} to aid
  comparison.} \label{eccentric}
\end{figure}

%
\section{Results}

There is a special case of the problem which can be solved using known
results. When $\beta=1$, the optimal balanced flow of fluid 1 must
mirror that in fluid 2. In particular, $\lambda=0$, $\Gamma$ is the
diameter $x=0$ {\em and} $u=0$ on $\Gamma$. The problems for either
fluid then decouple into single phase pressure-driven flow in a
`half'-cylinder (semicircular cross-section).  The flux is
$0.07438920$ in our non-dimensional units according to White's (1991)
equation (3-44). This provides an excellent test of the side-by-side
computations (see Table 1 which shows 3 significant figure
correspondence although there is really 5). Further checks are
available between the very different side-by-side and eccentric flow codes
(e.g. figure 4 where using $d$ as the abscissa shows at least $C^1$
continuity in $\max_{\lambda}Q$ at $\gamma =\pi$ or $d=0$ at
$\beta=5$).

\subsection{$max_{\lambda} Q$}

\begin{figure}
\centering \resizebox{\textwidth}{!}{\includegraphics{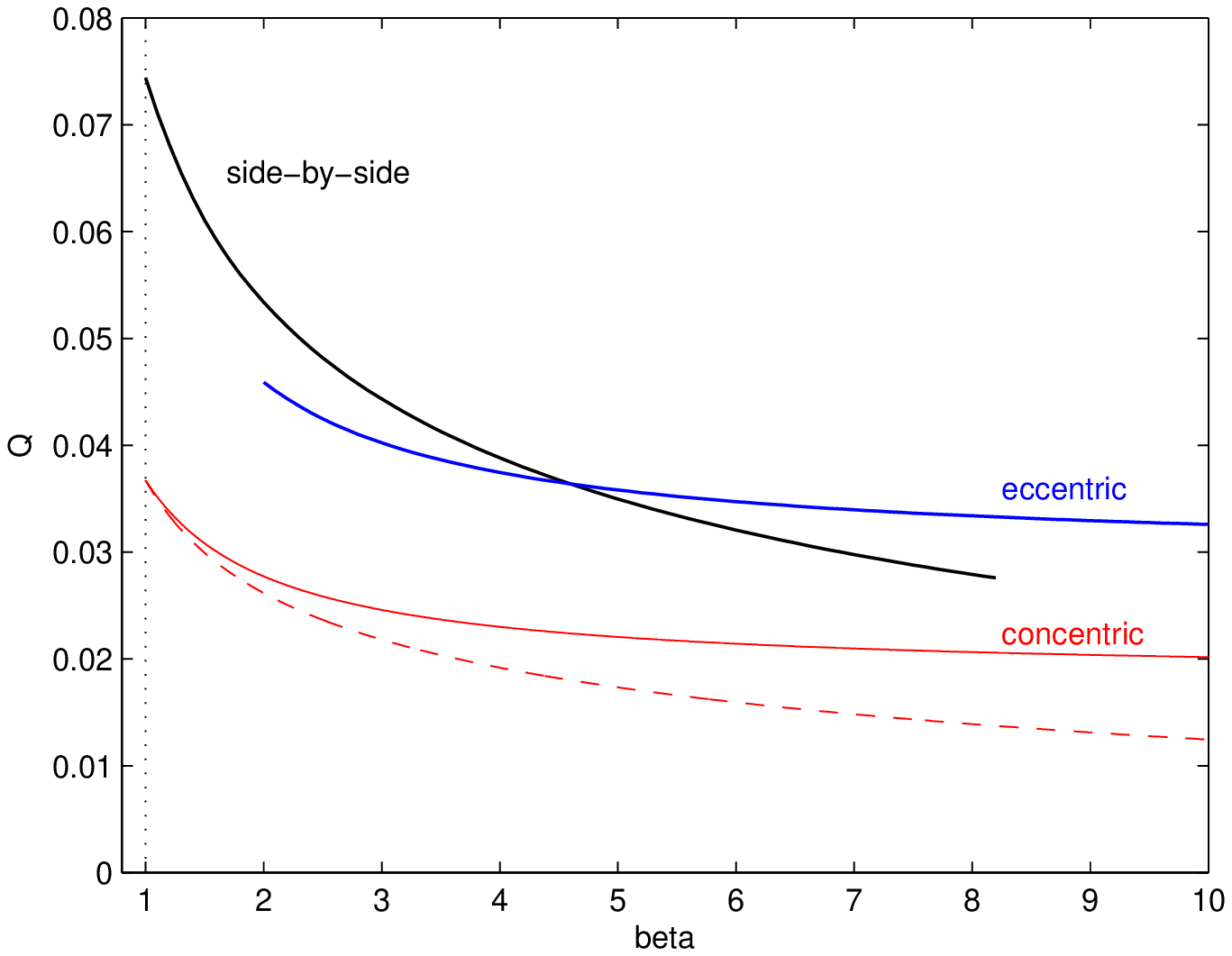}}
\caption{ ${\rm max}_{\lambda,\gamma} Q_s$ (left upper black curve),
  ${\rm max}_{\lambda,d} Q_e$ (right upper blue curve) and ${\rm
    max}_{\lambda} Q_c$ (lowest solid red curve) compared as a
  function of $\beta$. The side-by-side maximum disappears for $\beta
  \gtrsim 8.2$ and the eccentric solution only starts to have a
  maximum for $\beta \gtrsim 2$. The lowest dashed (red) curve
  corresponds to $max_{\lambda}\hat{Q}_c$, the global minimum of the more
  viscous fluid encapsulating the less viscous
  solution. }\label{Qbeta}
\end{figure}

The $\beta$ value chosen in figure \ref{intro} has been purposely
chosen to show the presence of flux maxima in the side-by-side
solutions and the ($d>0$ or more viscous fluid in the core) eccentric
solutions ($Q_e$). The complementary eccentric solutions with the less
viscous fluid in the core ($\hat{Q}_e$) always show monotonic
behaviour in which the flux decreases from the $\gamma=0$ side-by-side
value down to the concentric core-annular value of $\hat{Q}_c$
(leftmost point or most negative $d$). This uninteresting part of the
flux spectrum is suppressed in figure \ref{merge1} to focus on
$max_{\lambda} Q$ over $\gamma$ and $d>0$ for $\beta \in [2,10]$ over
which all the interesting behaviour occurs. At $\beta =1$, the
side-by-side solution with $\gamma=\pi/2$ and $\alpha=\pi/4$ supplies
the only flux maximum with both concentric core-annular solutions
being global minima as $Q_c=\hat{Q}_c$. At $\beta \approx 2$, a local
maximum starts to appear in the eccentric solutions with $d$ small and
positive (see figure \ref{merge1}). At $\beta \approx 4.60$, this
`eccentric' maximum becomes the global maximum with the `side-by-side'
local maximum disappearing by $\beta \approx 8.2$. Thereafter the sole
flux maximum is always an eccentric solution. Figures \ref{side} and
\ref{eccentric} show how the maxima change with $\beta$ including an
eccentric optimal flux solution at $\beta=10,000$. This confirms that
the optimal asymptotic solution has plug flow for the more viscous
core. Figure \ref{Qbeta} plots the maxima values as a function of
$\beta$ highlighting the cross-over point at $\beta \approx 4.60$ (see
also Tables 1 and 2). The concentric core-annular flux values for the
more viscous fluid in the core $Q_c$ and less viscous fluid in the
core $\hat{Q}_c$ are also shown as a local and global minima
respectively.

%
%

%
%
\begin{table}
\begin{center}
\begin{tabular}{@{}rrccccccc@{}}
  $\beta$             & \qquad  \quad &
  $\lambda$           & \qquad  \quad &
  $\gamma$            & \qquad  \quad &
  $\alpha$            & \qquad  \quad &
  $Q_s$ ($\times 10^{-2}$)    \\
          &&        &&       &&       &&         \\
  1       &&  0.00  && 1.57 && 0.785 && 7.44  \\
  1.5     && -0.06  && 1.52 && 0.810 && 6.11  \\
  2       && -0.10  && 1.50 && 0.812 && 5.34  \\
  2.5     && -0.13  && 1.48 && 0.814 && 4.82  \\
  3       && -0.15  && 1.46 && 0.813 && 4.43  \\
  3.5     && -0.18  && 1.46 && 0.809 && 4.13  \\
  4       && -0.19  && 1.47 && 0.786 && 3.88  \\
  4.5     && -0.21  && 1.47 && 0.779 && 3.67  \\
  5       && -0.22  && 1.48 && 0.761 && 3.50  \\
  6       && -0.25  && 1.52 && 0.721 && 3.21  \\
  7       && -0.27  && 1.58 && 0.668 && 2.98  \\
  8       && -0.28  && 1.69 && 0.585 && 2.79  \\
\\[6pt]
\end{tabular}
\end{center}
\caption{${\rm max}_{\lambda,\gamma}Q_s$ ($Q$ for the side-by-side
solution) as a function of $\beta$. The maximum is unique  global
for $\beta < 4.60$ and thereafter is a local maximum until it
vanishes for a $\beta$ $\approx8.2$.} \label{SideQ}
\end{table}

%
%
\begin{table}
\begin{center}
\begin{tabular}{@{}rrcclllll@{}}
  $\beta$             & \qquad  \quad &
  $\lambda$           & \qquad  \quad &
  $\sigma$            & \qquad  \quad &
  $R$            & \qquad  \quad &
  $Q_e$ ($\times 10^{-2}$)    \\
          &&        &&       &&       &&         \\
  2.5     && -0.140  && -0.393 && 0.594 && 4.25  \\
  3       && -0.155  && -0.390 && 0.590 && 4.02  \\
  3.5     && -0.168  && -0.387 && 0.584 && 3.86  \\
  4       && -0.180  && -0.387 && 0.582 && 3.74  \\
  4.5     && -0.188  && -0.386 && 0.578 && 3.65  \\
  5       && -0.198  && -0.386 && 0.574 && 3.58  \\
  6       && -0.209  && -0.385 && 0.570 && 3.47  \\
  7       && -0.211  && -0.382 && 0.568 && 3.40  \\
  8       && -0.222  && -0.383 && 0.565 && 3.34  \\
  9       && -0.225  && -0.383 && 0.564 && 3.29  \\
  10      && -0.226  && -0.381 && 0.563 && 3.26  \\
  15      && -0.246  && -0.383 && 0.555 && 3.15  \\
  20      && -0.245  && -0.381 && 0.556 && 3.10  \\
  50      && -0.260  && -0.381 && 0.549 && 3.01  \\
 100      && -0.260  && -0.380 && 0.551 && 2.98  \\
 200      && -0.263  && -0.381 && 0.550 && 2.96  \\
 500      && -0.263  && -0.3795 && 0.5485 && 2.9544  \\
1000      && -0.263  && -0.3794 && 0.5476 && 2.9514  \\
2000      && -0.263  && -0.3797 && 0.5473 && 2.9499  \\
5000      && -0.263  && -0.3795 && 0.5479 && 2.9490  \\
10000     && -0.263  && -0.3795 && 0.5479 && 2.9487  \\
          &&         &&        &&        &&         \\
$\infty$  && -0.263  && -0.3795 && 0.5480 && 2.9484  \\
\\[6pt]
\end{tabular}
\end{center}
\caption{${\rm max}_{\lambda,d}Q_e$ ($Q$ for the eccentric solution)
as a function of $\beta$. The maximum appears for $\beta \approx 2$,
is a local maximum for $2\lesssim \beta < 4.60$ and becomes a unique
  global maximum for $\beta > 4.60$.}
\label{EccQ}
\end{table}

%
%
%
\begin{figure}
\begin{center}
\setlength{\unitlength}{1cm}
\begin{picture}(14,14)
\psfrag{beta}{$\beta$}
\psfrag{Q}{$Q_{\infty}$}
\put(-0.5,0){\epsfig{figure=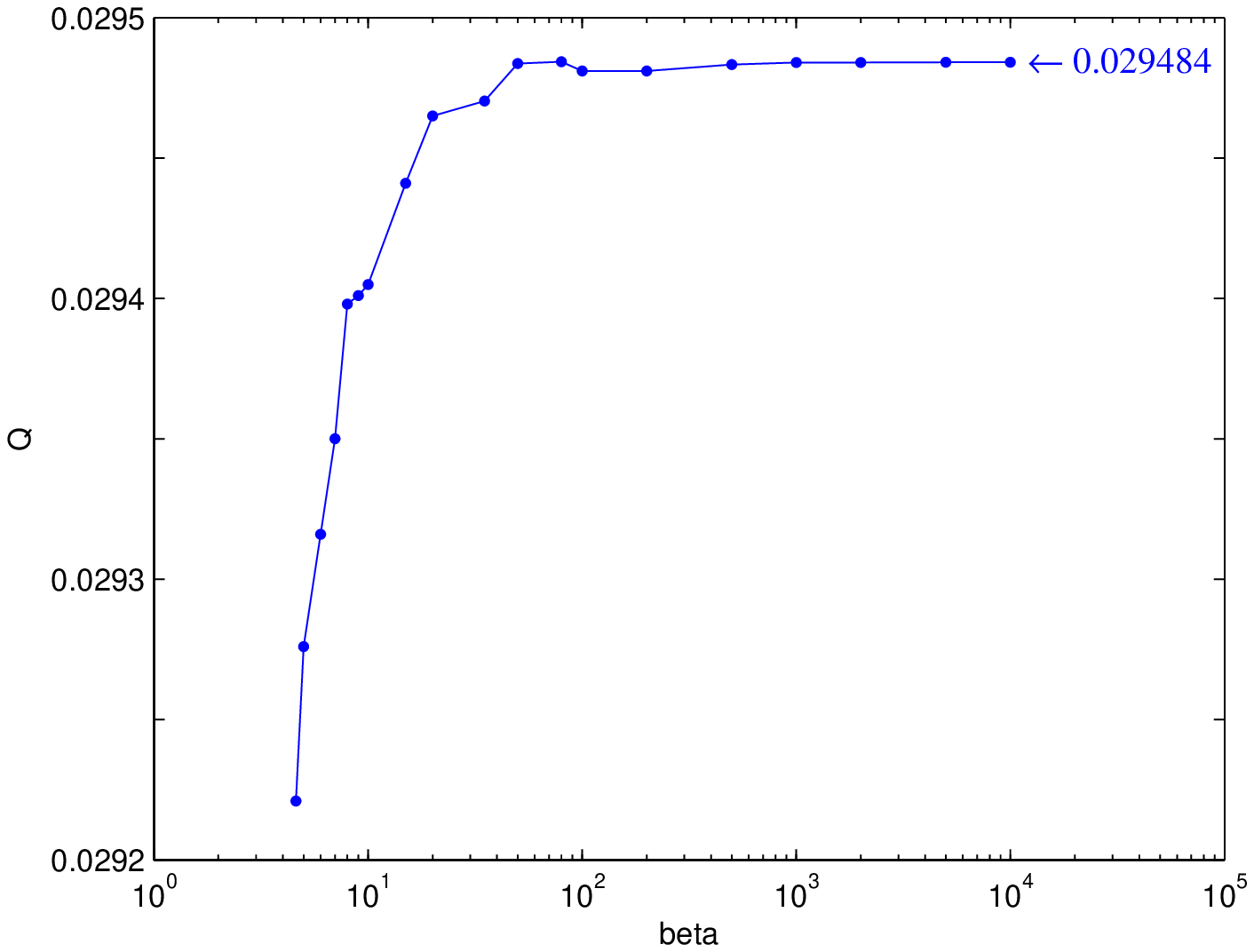,width=14cm,height=12cm}}
\psfrag{b}{$\beta$}
\psfrag{a}{$a_1$}
\put(3.75,1.5){\epsfig{figure=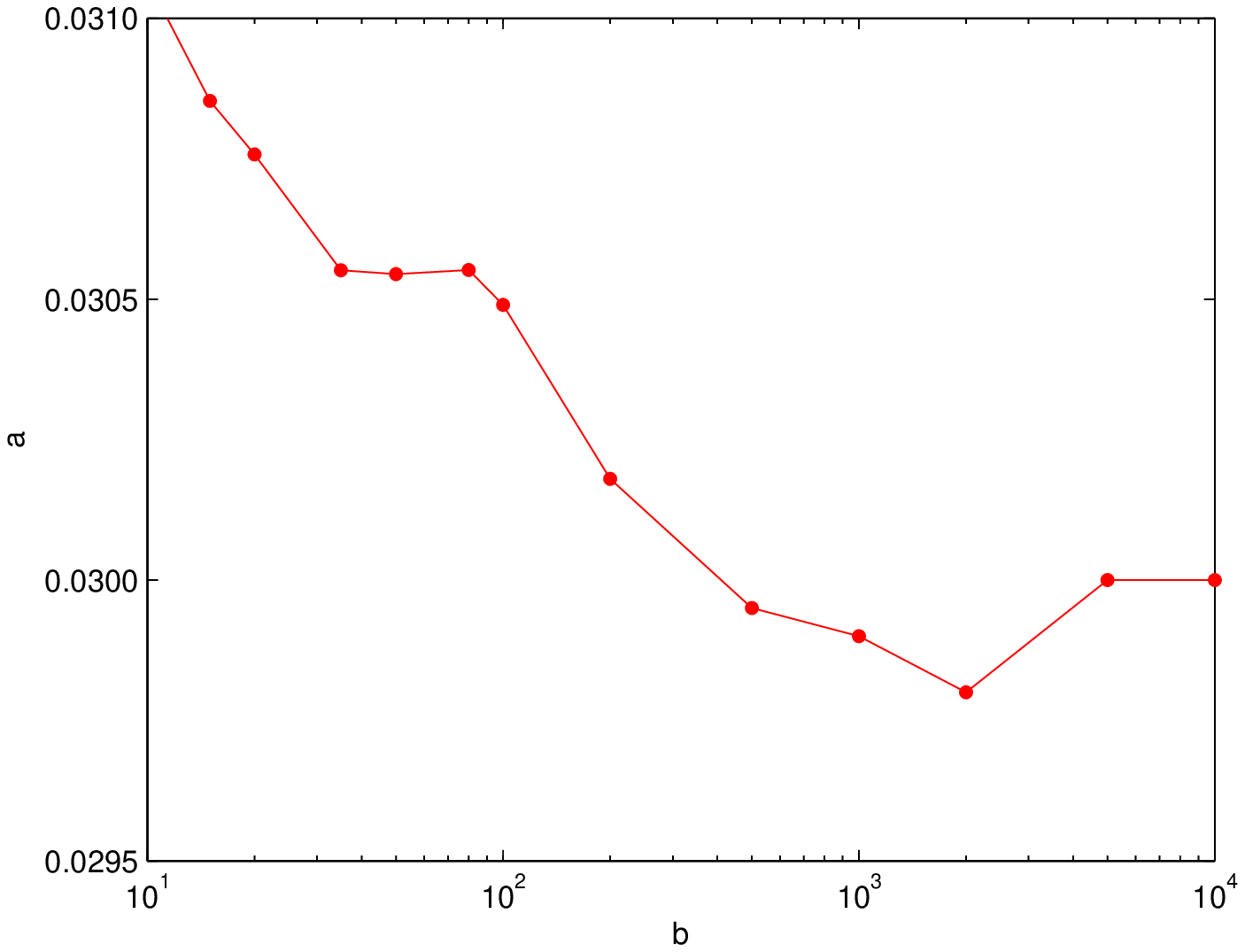,width=8.25cm,height=7.071cm}}
\end{picture}
\caption{The coefficients $Q_{\infty}$ and $a_1$
  (inset) as calculated using the expressions (\ref{asympt}) against $\beta$ where $\beta$ and the next smallest value of $\beta$ were used. }
\label{asymptotics}
 \end{center}
\end{figure}

\subsection{$max_{\lambda}Q$ for $\beta \rightarrow \infty$}

At large $\beta$, there is every reason to suspect that the maximal
flux possible possesses a simple expansion around its limiting value:
\beq
max_{\lambda,d}Q_e(\beta,\lambda;d)=
Q_{\infty}+\frac{a_1}{\beta}+\frac{a_2}{\beta^2}+\ldots
\eeq
The scalars $Q_{\infty}$ and $a_1$ can be estimated as follows
\beq
Q_{\infty} \approx \frac{ \beta_1 Q(\beta_1)-\beta_2
  Q(\beta_2)}{\beta_1-\beta_2}
\qquad
a_1 \approx \frac{\beta_1 \beta_2}{\beta_2-\beta_1}
\biggl[ Q(\beta_1)-Q(\beta_2) \biggr]
\label{asympt}
\eeq 
where $\beta_1$ and $\beta_2$ have suitably large values.  There is
good evidence that $Q_{\infty} \approx 2.9484 \times 10^{-2}$ and $a_1
\approx 3.00 \times 10^{-2}$ supporting the original assumption: see
figure \ref{asymptotics}. Another check on this value of $Q_{\infty}$
is available by artificially imposing plug flow in the core (e.g. see
the lower right solution in figure \ref{eccentric}). The matching
conditions at $\Gamma$ then simplify to just continuity $u_1=u_2$ and
the condition that the continuation of $u_1$ into $A_2$ has no
logarithmic singularities ($\oint_{\Gamma} {\bf dx.\nabla}u_1=0$)
which eliminates $\beta$ from the problem.  A straightforward search
over $\lambda$ and $\sigma$ then reveals the maximum of
$Q_{\infty}=2.94844 \times 10^{-2}$ at $\lambda=-0.263$
$\sigma=-0.3795$, $R=0.54798$ (and $d=1+\sigma-R=0.0725$).

\subsection{$Q(\beta,\lambda)$ for fixed $\lambda$}

\begin{figure}
\centering \setlength{\unitlength}{1cm}
\begin{picture}(14,14)
\put(-1,0){\includegraphics[width=16cm]{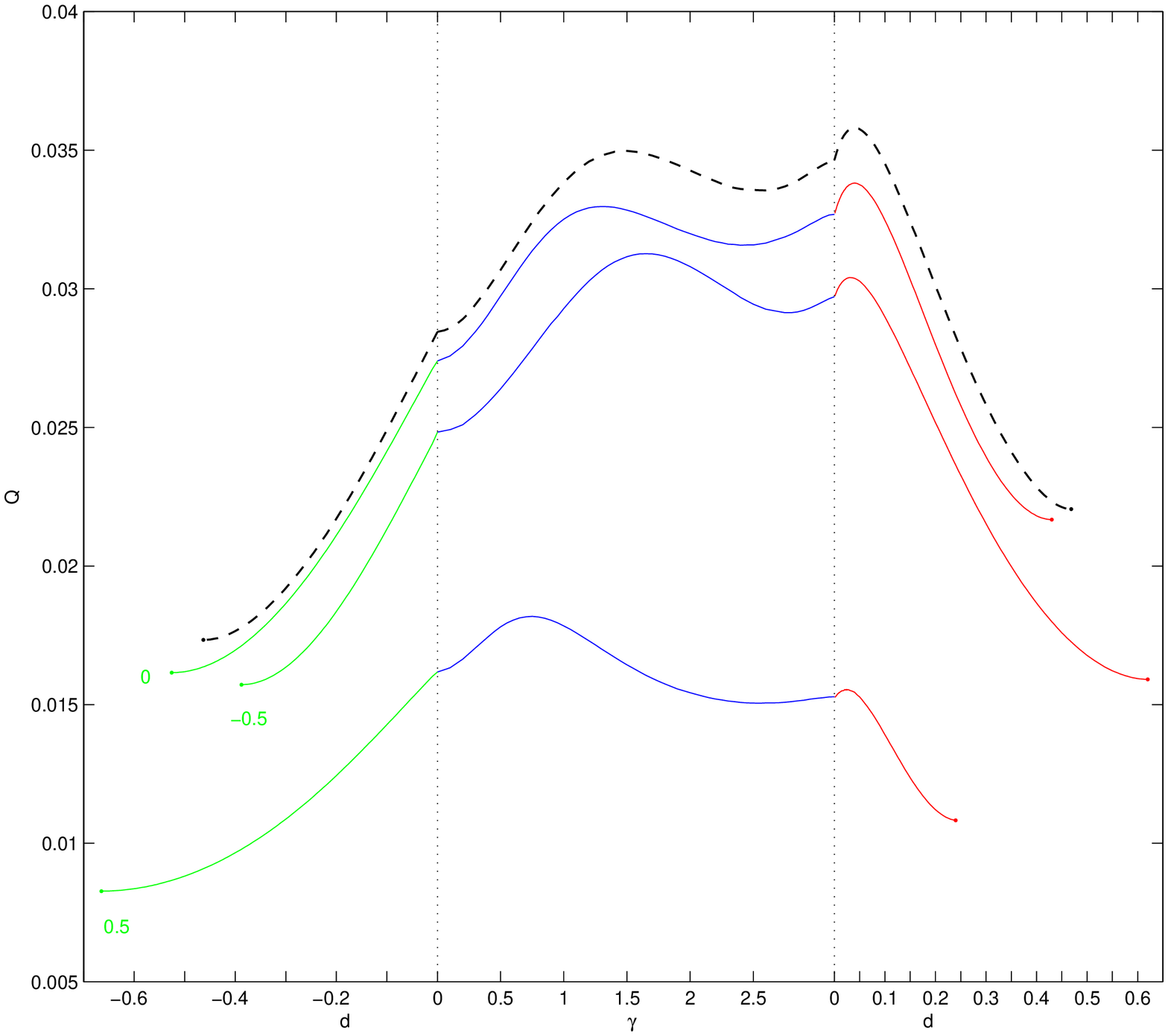}}
\end{picture}
\caption{The effect of fixing the pressure gradient $\lambda$ at
$-0.5$, $0$ and $0.5$ on $Q$ for $\beta=5$. The (black) dashed upper
envelope is the result of optimising over $\lambda$ as shown in
figure \ref{merge1}. } \label{merge2a}
\end{figure}

\begin{figure}
\centering 
\setlength{\unitlength}{1cm}
\begin{picture}(14,14)
\put(-1,0){\epsfig{figure=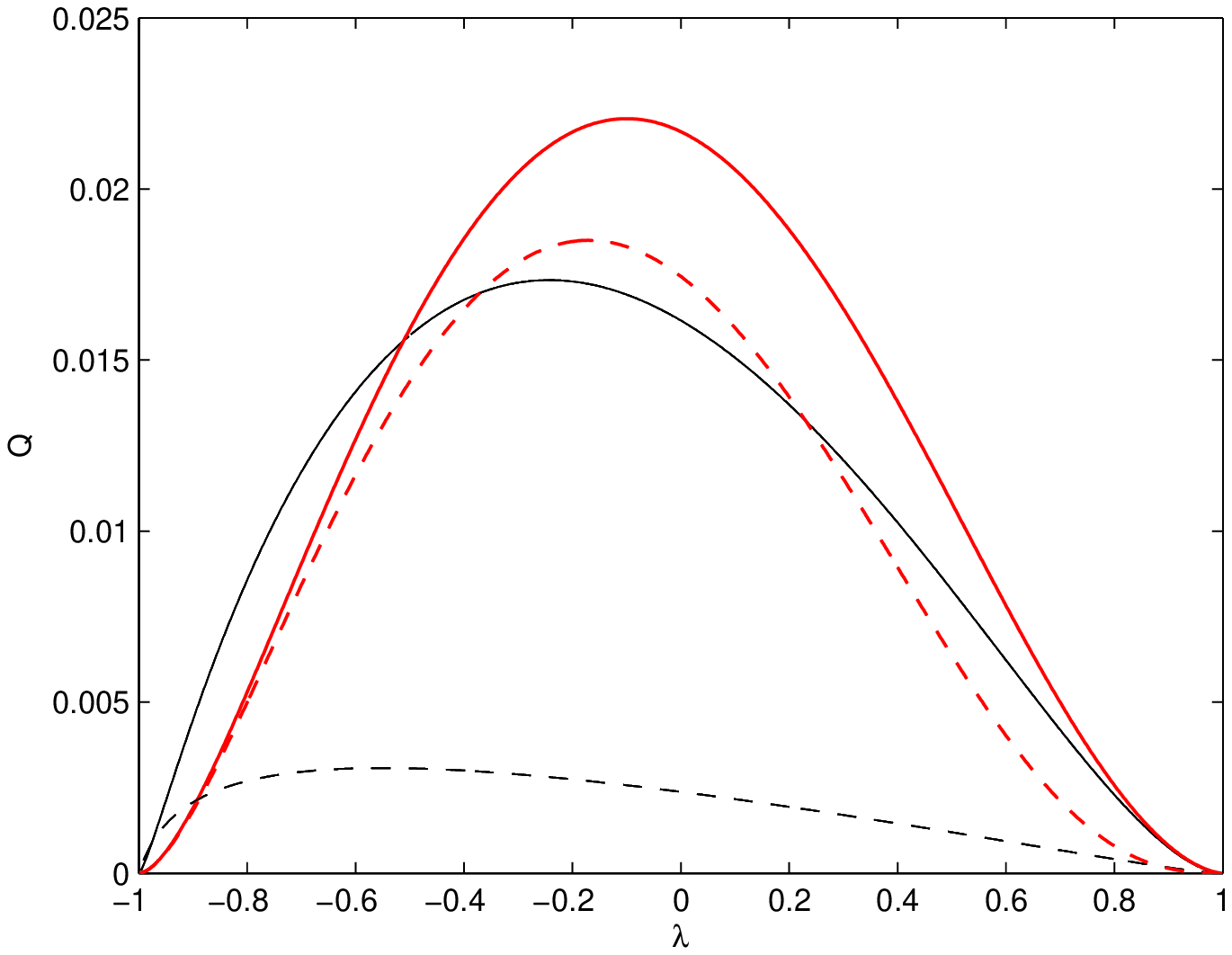,width=14cm,height=10cm}}
\end{picture}
\caption{The concentric core-annular fluxes $Q_c(\beta,\lambda)$ and
  $\hat{Q}_c(\beta,\lambda)$ plotted against $\lambda$ for $\beta=5$
  (thick solid red and thin solid black respectively) and $\beta=100$
  (thick dashed red and thin dashed black respectively). The crossing
  of the solid lines at $\approx -0.51$ is consistent with figure
  \ref{merge2a} where $Q_c \approx \hat{Q}_c$ at $\lambda=-0.5$. The
  dashed lines cross at $\lambda \approx -0.89$ for a ratio of
  100. } \label{comp}
\end{figure}

So far all the results shown have been optimised over the pressure
gradient $\lambda$. The presumption is that, in the absence of any
explicitly imposed gradient, the flow sets up its own to maximum the
volumetric exchange. Figure \ref{merge2a} shows the effect of fixing
$\lambda$ on the flux profile at $\beta=5$. The same general trends
emerge with one important additional feature highlighted by the
$\lambda=-0.5$ curve. Here $\hat{Q}_c$ (leftmost point) is
approximately the same as $Q_c$ (rightmost point). Figure \ref{comp}
plots the two core-annular flux functions $Q_c$ and $\hat{Q}_c$
against $\lambda$ to show that the less-viscous core solution flux
$\hat{Q}_c$ actually exceeds the more-viscous core solution flux $Q_c$
for $\lambda \lesssim -0.51$ at $\beta=5$. This threshold pressure
gradient montonically decreases as $\beta$ increases to, for example,
$\approx -0.89$ at $\beta=100$ (recall $-1 < \lambda < 1$): see figure
\ref{comp}. Since a $\lambda$ value of -1 translates into a pressure
gradient which hydrostatically maintains the denser fluid, the
conclusion is that the less-viscous-fluid-in-the-core concentric
solution is favoured over its complement for large enough pressure
gradients.

%
%

%
%

%
%
\section{Discussion}

This paper has considered the steady, coaxial flow of two immiscible
fluids of different densities and viscosities in a straight vertical
cylindrical tube such that their volumetric fluxes balance. Under mild
assumptions concerning the interface between the two fluids, the main
conclusion is that the flow which optimises the volumetric flux over
all possible pressure gradients is always asymmetric. In particular,
for viscosity ratios $\lesssim 4.60$ the optimal flow is a
side-by-side solution in which each fluid makes contact with a side of
the tube and otherwise is an {\em eccentric} core-annular solution
with the more viscous fluid encapsulated by the less viscous
fluid. (In fact, in this latter case, the eccentricity is so marked,
that it could look like a side-by-side solution from one direction to
the unwary.)  The axisymmetric (concentric) core-annular solution in
which one fluid encircles the other is surprisingly either a local or
global minimiser of the flux. The clear conclusion is that displacing
the core of such a flow to one side increases the flux by allowing the
outer fluid to `bulge' through the larger gap. This generalises the
equivalent observation made for the flow of a single fluid through an
eccentric annulus duct (see figure 3-8 on page 127 of White 1991).

The fact that the principle of maximum flux predicts a side-by-side
solution at low viscosity ratios does find support in the work of
Arakeri et al (2000) and the experiments at Bristol (Beckett et
al. 2009). However, Huppert \& Hallworth (2007) never mention seeing a
side-by-side solution during their low-viscosity-ratio experiments,
instead reporting only a steady concentric core-annular flow. More
intriguing, however, is that in this core-annular solution, both
Huppert \& Hallworth (2007) and Beckett et al. (2009) invariably see
the lower (less dense) fluid {\em rising} along the axis. On the basis
that less dense fluids generically are less viscous too, this implies
that the less viscous fluid is typically at the core of these observed
flows. From the flux perspective, the results presented here show that
this globally {\em minimises} the flux over all possible pressure
gradients! This apparent contradiction is ameliorated somewhat if the
pressure gradient set up (or imposed) is towards the maximum possible
for exchange (e.g. see figure \ref{comp}), but nevertheless the
core-annular solution still remains a local flux minimiser.  The
principle of {\em minimum} flux (and, coincidentally, minimum
dissipation) then appears more useful at large viscosity ratios.

The proper route to resolving this conundrum, of course, is careful
consideration of the initial value problem and the stability of the
evolving solution to the small disturbances always present. A first
step in this direction would be to study the Rayleigh-Taylor
instability problem in a cylindrical tube where a fluid of density
$\rho_1$ and viscosity $\mu_1$ fills the half cylinder $z>0$ and a
fluid of density $\rho_2 < \rho_1$ and viscosity $\mu_2$ occupies
$z<0$. Establishing which interfacial deformation mode (axisymmetric
or asymmetric) has the largest growth rate as a function of all the
parameters present would surely go some way in predicting which type
of flow is initiated. However, even this calculation doesn't seem to
have been done yet although Batchelor \& Nitsche (1993) come close.

In conclusion, it should be clear that there are some interesting
issues surrounding the exchange flow of two fluids in a vertical tube.
Even the steady immiscible problem displays an intriguing degeneracy
of solution. Focussing on an {\em ad hoc} principle of maximum (or
minium) flux unfortunately looks to be too simplistic despite its
appealing rationale and apparent success in an associated
context. This means that there is no avoiding a more formal
stability-based approach to explain what is seen in experiments.

\section{Acknowledgements}

This study was stimulated by ongoing experimental work carried out by
the Volcanology group in Earth Sciences at Bristol University (Frances
Beckett, Fred Witham, Jerry Phillips and Heidy Mader) with whom I have
enjoyed many stimulating discussions. I would also like to thank Carl
Dettmann for a reassuring discussion on singular integrals and Diki
Porter for sharing his expertise on solving Laplace's equation in
complex geometries.

%
%

\begin{appendix}

\section{Side-by-side solutions}

The geometry of the side-by-side solution is shown in figure
\ref{zdiagram} to be defined by two parameters: $\gamma$, the (upper)
intercept latitude of $\Gamma$ with the duct wall, and $2\alpha$, the
angle between $\Gamma$ and duct wall.  The coupled Poisson problems
(\ref{prob1})-(\ref{bcs}) become two Laplace problems by separating
off simple inhomogeneous parts as follows
\beq
u_1^*=\Phi_1+\frac{\lambda+1}{4}      (\,x^2+y^2-1\,), \qquad
u_2^*=\Phi_2+\frac{\lambda-1}{4 \beta}(\,x^2+y^2-1\,)
\eeq
which have been designed to leave the boundary conditions on the duct
wall undisturbed. The functions $\Phi_1$ and $\Phi_2$ then
satisfy
\begin{eqnarray}
       \nabla^2 \Phi_1 &=& 0\quad {\rm in} \quad A_1,\\
       \nabla^2 \Phi_2 &=& 0\quad {\rm in} \quad A_2,
\end{eqnarray}
with boundary conditions
\begin{eqnarray}
\Phi_1 =0 && {\rm on} \quad x+iy =e^{i\psi} \quad
-\gamma \leq \psi \leq \gamma \label{bc1}\\
\Phi_2=0
&& {\rm on} \quad x+iy =e^{i\psi} \quad \gamma \leq \psi \leq 2 \pi-\gamma
\label{bc2}
\end{eqnarray}
\beq
\biggl.
\begin{array}{rcr}
\Phi_1-\Phi_2\,\,\, &=& (\frac{\lambda+1}{4}-\frac{\lambda-1}{4 \beta})
(\,1-x^2-y^2\,) \\
2\frac{\partial}{\partial n} (\Phi_1-\beta \Phi_2)
&=&\frac{\partial}{\partial n}
(\,1-x^2-y^2\,)
\end{array} \biggr\}
\quad {\rm on} \quad x+iy \in \Gamma\\
\label{conditions}
\eeq
Here the interface curve $\Gamma:=\{\, z\,|\, z=x+iy=\sigma+R e^{i
  \theta}\,;\,|\theta| \leq |\theta_{max}:=\gamma-2\alpha|\, \}$ where
\beq
\sigma:=-\frac{\sin 2 \alpha}{\sin \theta_{max}} \quad \& \quad
R:=\frac{\sin \gamma}{\sin \theta_{max}}
\label{defnR}
\eeq
are formulae for the centre $(x,y)=(\sigma,0)$ and radius of curvature
respectively valid for any pair $0\leq 2\alpha,\gamma \leq \pi$. (The
singular case $\gamma=2 \alpha$ where $R \rightarrow \infty$ so that
$\Gamma$ is a straight line cannot be formally handled but is never a
practical problem.)
%
%

%
%

\begin{figure}
\hspace{-0.25cm}
   \begin{tabular}{cc}
      \epsfig{figure=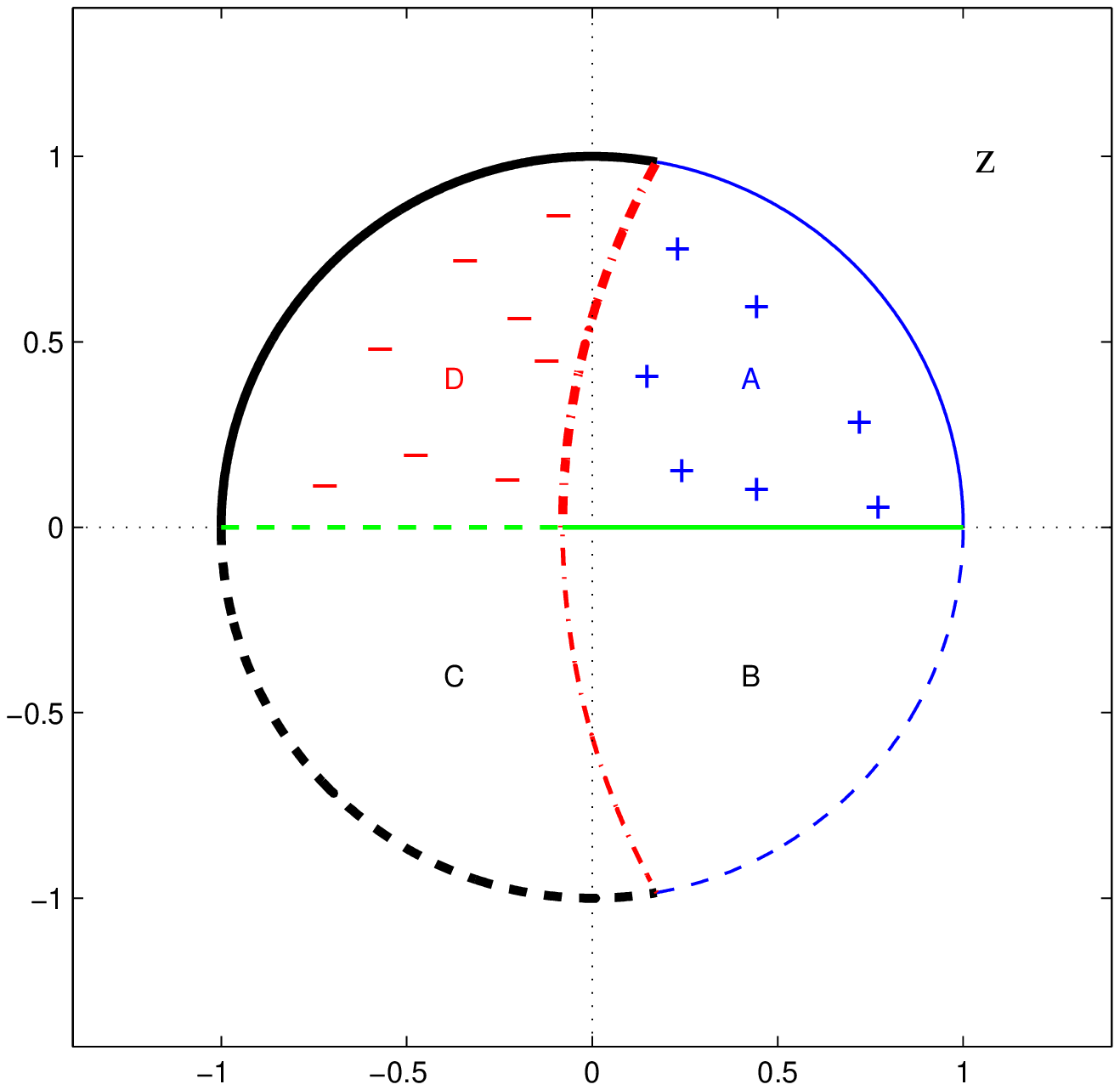, scale=0.5} &
      \epsfig{figure=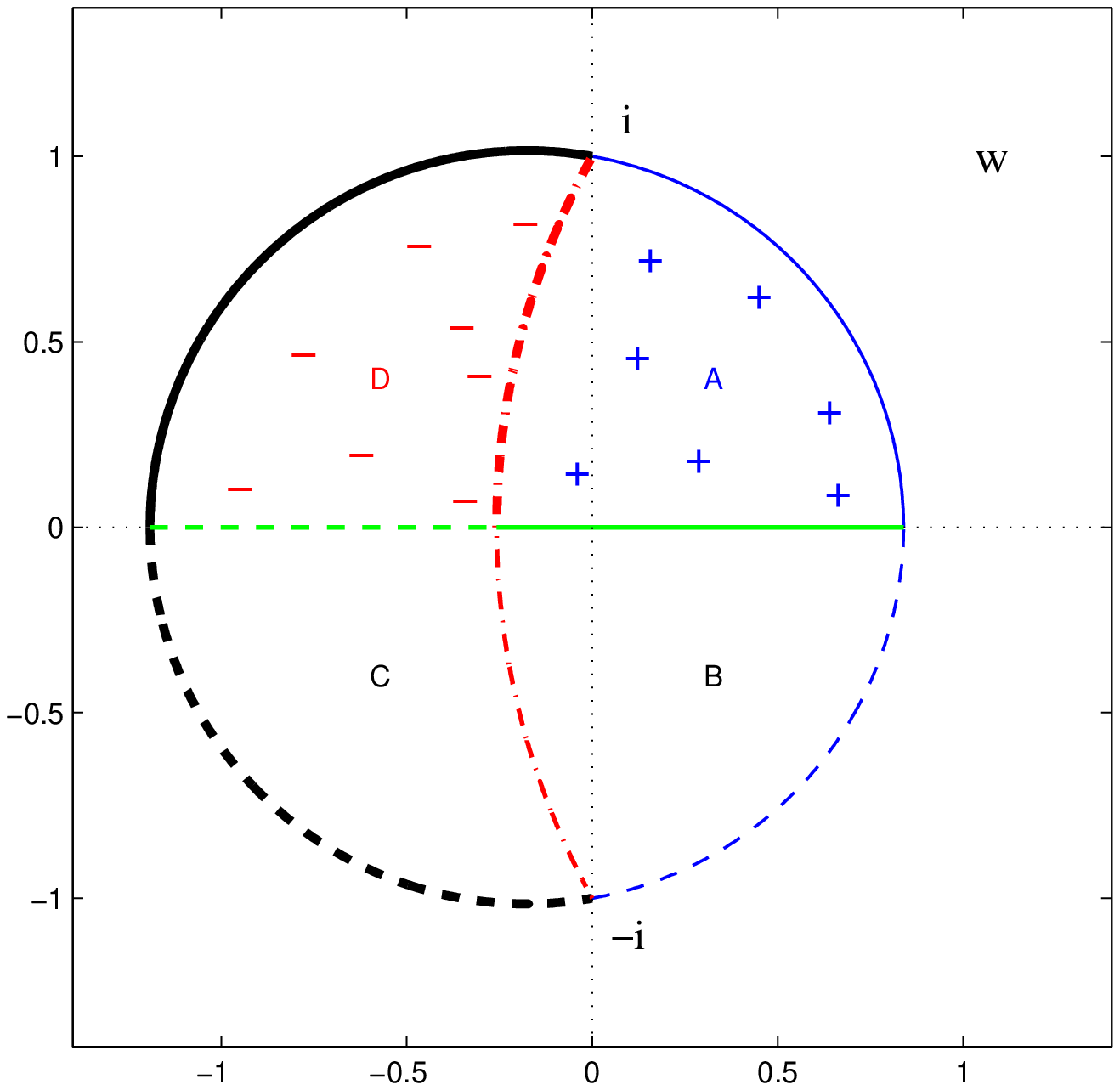, scale=0.5} \\
      \epsfig{figure=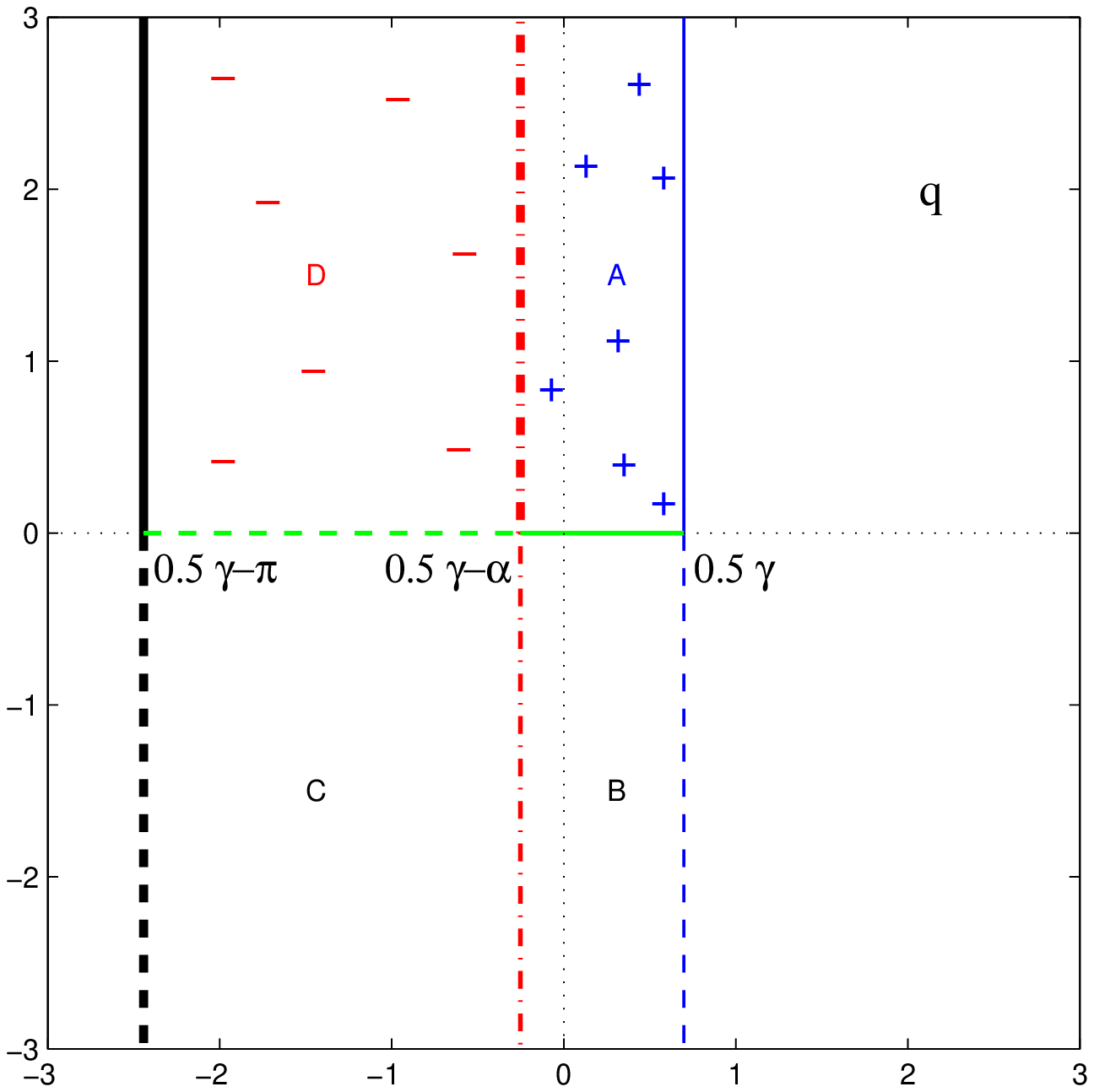, scale=0.5} &
      \epsfig{figure=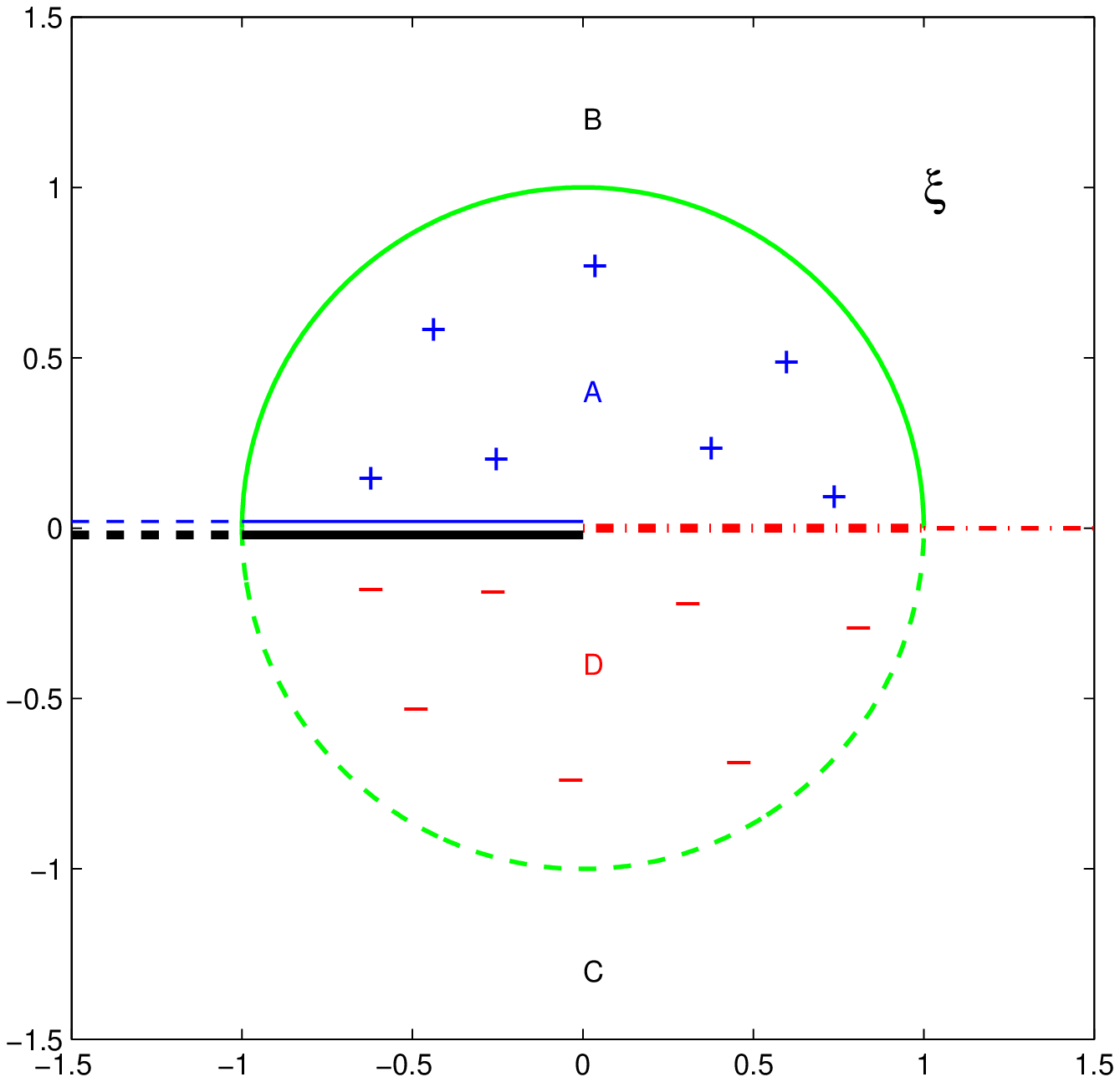, scale=0.5}
   \end{tabular}
   \caption{The composite conformal mapping from $z$ to $\xi$. The
     interface curve $\Gamma$ meets the duct wall at $e^{\pm i
       \gamma}$ in the $z$-plane, which are moved to $\pm i$ in the
     $w$-plane, and then to $\pm i \infty$ in the $q$-plane. Two
     separate transformations then map the strip $\half \gamma-\alpha
     \leq Re(q) \leq \half \gamma$ onto the upper half plane
     and the strip $\half \gamma-\pi \leq
     Re(q) \leq \half \gamma-\alpha$ onto the lower half
     plane. In both cases, $Im(q)>0$ in mapped inside
     the appropriate unit semicircle in the $\xi$-plane. }
\label{mapping}
\end{figure}

The solution strategy is to transform regions $A_1$ and $A_2$ into
the upper and lower half planes respectively via conformal
tranformations where an explicit solution can then be deduced by
Poisson's integral formula. Three simple transformations prove
sufficient, the first is
\beq
w:=\frac{z-\cos \gamma}{\sin \gamma}
\eeq
which rescales the duct so that its  radius becomes  $1/\sin \gamma$ and
the 2 contact points of $\Gamma$ with the duct wall $e^{\pm i \gamma}$
move to $\pm i$ (other noteworthy images are:
$0 \rightarrow  -\cot \gamma$,
$1 \rightarrow   \sin \gamma/(1+\cos \gamma)$,
$-1 \rightarrow -\sin \gamma/(1-\cos \gamma)$
; see Figure 2).
A second transformation
\beq
q=r+is:= \tan^{-1} \,w \qquad \qquad r,s \in \Re e
\label{trans2}
\eeq
converts  all the  circular arcs into straight lines
parallel to the imaginary axis in the complex $q$ plane. To see this,
consider the transformation in reverse
\beq
w=\tan q =-i \frac{e^{2 i q}-1}{e^{2 i q}+1}
\eeq
and decompose this transformation into its 3 components. A strip
$\half \gamma-\alpha \leq r \leq \half \gamma$ with $0<\alpha \leq
\pi$ is rotated through $\pi/2$ by $q \rightarrow iq$.  Doubling and
exponentiating $q\rightarrow iq \rightarrow e^{2iq}$ then tranforms
the strip into the interior of a wedge centred at the origin with
sides of argument $\gamma - 2 \alpha$ and $\gamma$. Finally the
M\"{o}bius transformation $q\rightarrow iq \rightarrow e^{2iq}
\rightarrow -i(e^{2iq}-1)/(e^{2iq}+1)$ converts the wedge sides into
circular arcs joining the points $w=\pm i$ and the wedge interior into
a circular lune with angle $2 \alpha$ (see Figure 2 and pages 205-207
of Marushevich 1965). The conformal transformation (\ref{trans2}) is
undoubtedly not the only one which would do the job (e.g. Vlasov 1986)
but is particularly nice since it can used to treat both `lunes'
together: $A_1$ maps to the strip $\half \gamma-\alpha \leq r \leq
\half \gamma$ and $A_2$ maps into the strip $\half \gamma-\half \pi
\leq r \leq \half \gamma -\alpha$ in the $q$-plane. The intersection
$A_1 \cap A_2=\Gamma$ is then the line $\Re e(q)=r=\half
\gamma-\alpha$.

The final transformation does, however, need  tailoring to each domain
separately as follows
\begin{eqnarray}
\xi=\xi_1(q) &:=& e^{i \pi(2q-\gamma+2\alpha)/2\alpha} \qquad \hspace{0.6cm}
q \, \in \, A_1 \\
\xi=\xi_2(q) &:=& e^{i \pi(2q-\gamma+2\alpha)/(\pi-2 \alpha)}
\qquad q \, \in \, A_2
\end{eqnarray}
so that the final composition transformations are
\begin{eqnarray}
\xi_1(z) &:=& \exp \biggl(\frac{i \pi}{\alpha} [\tan^{-1}w(z)-\half \gamma+\alpha ]\biggr),
\\
\xi_2(z) &:=& \exp \biggl(\frac{i \pi}{\half \pi-\alpha} [\tan^{-1}w(z)-\half \gamma+\alpha ]\biggr)
\end{eqnarray}
where
\begin{equation}
w(z)=\frac{z-\cos \gamma}{\sin \gamma}=\frac{e^{i \theta}-\cos(2 \alpha-\gamma)}{\sin(\gamma-2\alpha)}.
\end{equation}
The image of $A_1/A_2$ is designed as the upper/lower half $\xi$-plane
and $\Gamma$ remains a shared boundary (see Figure 2). If we define
$\xi=\zeta+i\eta$ and $\overline{\Phi}_{i}(\,\zeta(x,y),\eta(x,y)\,):=\Phi_{i}(x,y)$ ($i=1,2$), the
solutions for $\overline{\Phi}_1$ and $\overline{\Phi}_2$ are then
available via Poisson's integral formula for the half plane
\begin{eqnarray}
\overline{\Phi}_1(\zeta,\eta)
&=&
\quad\frac{1}{\pi}\int^{\infty}_{-\infty} \frac{\eta \overline{\Phi}_1(t,0)}
{(\zeta-t)^2+\eta^2} \, d t \label{P1} \\
\overline{\Phi}_2(\zeta,\eta)
&=&
-\frac{1}{\pi}\int^{\infty}_{-\infty} \frac{\eta \overline{\Phi}_2(t,0)}
{(\zeta-t)^2+\eta^2} \, d t  \label{P2}
\end{eqnarray}
The conditions (\ref{bc1}) and (\ref{bc2}) indicate that
$\overline{\Phi}_1$ and $\overline{\Phi}_2$ are only non-zero on the
image of $\Gamma$ which is the positive real axis ($t \geq 0$) in the
$\xi$-plane. The problem now boils down to determining the function
$f(z):=u^*_1=u^*_2$ on $\Gamma$ such that the stress matching
condition (see (\ref{bcs}) on $\Gamma$ holds. Applying this condition
is slightly non-trivial because the integrals (\ref{P1}) and
(\ref{P2}) are formally singular for $\xi=\zeta+i \eta$ on
$\Gamma$. They have well-defined (Cauchy principal) values by
continuity with surrounding values of $\xi$ but taking normal
derivatives of these integrals and subsequently computing them,
nevertheless, requires due care. Consider the normal ($\eta$)
derivative of $\overline{\Phi}_1$ on $\Gamma$ ($\eta=0$), for
example. It is straightforward to show
\beq
\overline{\Phi}_{1,\eta}(\zeta,\eta)=
\quad\frac{1}{\pi}\int^{\infty}_{-\infty} \overline{\Phi}_1(t,0)
\frac{\partial}{\partial t}\left[
  \frac{\zeta-t}{(\zeta-t)^2+\eta^2}\right] \, d t.
\eeq
and, after integration by parts, then
\beq
\overline{\Phi}_{1,\eta}(\zeta,0)= \frac{1}{\pi}\int^{\infty}_{-\infty}
\frac{\overline{\Phi}_{1,\zeta}(t,0)}{t-\zeta} \,
d t=
\frac{1}{\pi}\int^{\infty}_{-\infty}
\frac{\overline{\Phi}_{1,\zeta}(t,0)-
      \overline{\Phi}_{1,\zeta}(\zeta,0)}
    {t-\zeta} \,dt
\label{regularisation}
\eeq
since the Cauchy principal value of
$\int^{\infty}_{-\infty}1/(t-\zeta) dt$ is zero. The last integral on
the right hand side of (\ref{regularisation}) is now regular. The
symmetry of the velocity fields under $y \rightarrow -y$ in the
$z$-plane can then be invoked to make the integration range
finite.  This reflectional symmetry carries over to the $ \xi$-plane
as the symmetry $\overline{\Phi}_i(1/t,0)=\overline{\Phi}_i(t,0)$
($i=1,2$) allowing, for example, (\ref{P1}) to be simplified to
\beq
\overline{\Phi}_1(\zeta,\eta)= \frac{\eta}{\pi}\int^{1}_{0}
\overline{\Phi}_1(t,0)
\biggl[ \frac{1}{(\zeta-t)^2+\eta^2}
+\frac{1}{(t\zeta-1)^2+t^2 \eta^2} \biggr]\, d t.
\label{P1better}
\eeq
and (\ref{regularisation}) to
\beq
\overline{\Phi}_{1,\eta}(\zeta,0)=
\frac{1}{\pi}\int^{1}_{0} \biggl[\,
\frac{ \overline{\Phi}_{1,\zeta}(t,0)
      -\overline{\Phi}_{1,\zeta}(\zeta,0)}{t-\zeta}
+
\frac{t \overline{\Phi}_{1,\zeta}(t,0)}{\zeta t-1}
\biggr]
\,dt
+ \frac{\overline{\Phi}_{1,\zeta}(\zeta,0)}{\pi}
\log \left( \frac{1-\zeta}{\zeta} \right).
\label{dP1better}
\eeq
These are the integral representations (along with the equivalent ones
for $\overline{\Phi}_2$) used to impose the matching
conditions and calculate the flow solution.

In the matching process, the first step in determining $f$ is to
construct a global representation, $f(\theta)=\sum^N_{n=1} c_n \Psi_n(\theta)$,
using $\theta$ to parametrise  $\Gamma$, $c_n$ as the
expansion constants and the basis functions
\beq
\Psi_n(\theta):=T_{2n}(\theta/\theta_{max})
             -T_{2n-2}(\theta/\theta_{max}).
\eeq
These are defined in terms of Chebyshev polynomials
$T_n(\theta):=\cos(n \cos^{-1} \theta)$ with each designed to
mirror the properties of $f$: $f(\pm \theta_{max})=0$ and
$df/d\theta|_{\theta=0}=0$ by the $y-$reflectional symmetry. This symmetry
also means that the matching condition needs only to be applied (via
collocation at the $N$ positive zeros of $T_{2N+1}$) over the upper
half of $\Gamma$. It is tempting to carry out this procedure directly
in the $\xi-$plane using the representation (\ref{dP1better}) and the
sister integral for $\overline{\Phi}_{2,\eta}$. However, this proves
inaccurate because both have an integrable singularity at $t=0$
($\theta=\pm \theta_{max}$). This causes loss of accuracy through two
separate effects: a) the integrand has a singular derivative at $t=0$
so numerical quadrature is inefficient and b) the collocation points
sparsely populate the neighbourhood of $t=0$ at extreme choices of
$\alpha$ ($\rightarrow 0$ or $\pi/2$) so the matching is not well
imposed and convergence fails short of usual spectral (exponential)
accuracy. Instead, the integral representations must be transformed to
the physical $z-$plane and matching carried out there.

The velocity profile along $\Gamma$ is always smooth and typically
only $N=20$ or $30$ is needed to see spectral drop off of 4-5 orders
of magnitude.  The limits $\alpha \rightarrow \pi/2$ and $\alpha \rightarrow 0$, however, have to be treated carefully. 
For example, when $\alpha \geq 0.2$ ($\approx 10^o$)
only a 100-panel Simpson quadrature is needed to accurately calculate
the integrals along $\Gamma$ but this must be increased dramatically
as $\alpha \rightarrow 0$ due to the extreme behaviour of the
$z=z(\xi)$ transformation in this limit (e.g. $10^4$ panels proved
sufficient for $\alpha=O(0.001)$). Once the solution is obtained, the
fluxes $Q_1$ and $Q_2$ are calculated using Simpson's rule with
typically $20-40$ panels. This is the most costly part of the process
as essentially a triple integral is being evaluated. Simple bisection
in $\alpha$ is used to find a `balanced' flux state where $Q_1+Q_2=0$
for given $\beta$, $\lambda$ and $\gamma$.

As a final comment, it's worth remarking that the transformation
$q=q(z)$ (see the third subplot in figure \ref{mapping}) achieves a
separation of variables in the problem (the boundaries are contours of
constant $r=Re(q)$)\footnote{The transformation $q=q(z)$ is
  essentially a transformation to bipolar coordinates}. A solution
could therefore be developed by separation of variables after a
Fourier transform (in $s$) is taken of the inhomogeneity in the
matching condition. The full procedure, however, boils down to
essentially the same problem of evaluating a triple integral albeit in
this case the innermost one for $u$ is an inverse Fourier transform
and hence over a semi-infinite interval.


\section{Eccentric solutions}

The `eccentric' solution has one fluid completely encapsulated by the
other.  For sake of argument, we describe the solution strategy for
$A_2$ in $A_1$. The radius $R$ and centre $(\sigma,0)$ (with $\sigma <
0$) define the geometry uniquely up to an arbitrary rotation around
the duct axis and any reflection about a diameter neither of which, of
course, affect the flux. The interface curve $\Gamma$ is then
\beq
\Gamma:=\{\,z\,|\, z=x+iy=\sigma+R e^{i \theta}\,;\, -\pi < \theta \leq \pi\,\}
\eeq
which smoothly connects to the formula for $\Gamma$ in the
side-by-side solution (formally, $R$ is +/$-$ve if $\Gamma$ is convex/concave
as viewed from $x=-\infty$: see the definition (\ref{defnR}) ).  The
problem (\ref{prob1})$-$(\ref{bcs}) is solved by conformally mapping the
geometry of eccentric circles into one of concentric circles using a
bilinear transformation $\xi=\xi(z)$. This is constructed by selecting
a common pair of real inverse points $(\kappa,0)$ and $(\nu,0)$ for
$\Gamma$ and the duct wall $|z|=1$ (so $ |\kappa \nu|=1$ and
$|\kappa-\sigma||\nu-\sigma|=R^2$) which ensures that the transformation
\beq
\xi:=\frac{z-\kappa}{z-\nu}
\eeq
maps the two circles $|z|=1$ and $\Gamma$ into
concentric circles of radii (respectively)
\beq
\varpi_1 :=\frac{1-\kappa}{1-\nu} \quad \& \quad
\varpi_2 :=\frac{R+\sigma-\kappa}{R+\sigma-\nu}
\eeq
where
\beq
\left.
\begin{array}{c}
\kappa\\
\nu
\end{array}
\right\}
:=\frac{\pm(1+\sigma^2-R^2)- \sqrt{(1+\sigma^2-R^2)^2-4
    \sigma^2}}{2 \sigma}
\eeq
(so $\nu<-1$). In the $\xi=\varpi e^{i \phi}$
plane, the solution is found standardly using the expansions
\begin{eqnarray}
&&u_1=\sum^N_{n=1} A_n
\left( \varpi^n -\frac{\varpi_1^{2n}}{\varpi^{n}} \right)
\cos n \phi+A_0\log(\varpi/\varpi_1)
+\frac{\lambda+1}{4}(|z|^2-1), \nonumber \\
&&u_2=\sum^N_{n=0} B_n \varpi^n \cos n \phi+\frac{\lambda-1}{4
    \beta}|z|^2 \nonumber
\end{eqnarray}
which incorporate the boundary condition at $|z|=1$
($\varpi=\varpi_1$) and the $y-$symmetry ($\phi \rightarrow -\phi$) of
the problem.  The Fourier series in $\phi$ of $|z|=|\nu
\xi-\kappa|/|\xi -1|$ and $\partial |z|/\partial \varpi$ on $\Gamma$
need to be evaluated to apply the remaining matching conditions. This
is done routinely using Simpson's rule with 200 panels when
$N=100$. In the limiting situations of $\sigma-R \rightarrow -1$
($\Gamma$ approaching the duct wall) and $\sigma \rightarrow 0$
(approaching concentricity), these numbers are doubled to 400 and
$N=200$ to maintain at worst $10^{-10}$ least square error in either
matching condition. Calculation of the fluxes in $A_1$ and $A_2$ is
again by 2D Simpson's rule using 100-200 panels per direction and
simple bisection is used in $R$ used to identify where $Q_1+Q_2=0$ for
given $\beta$, $\lambda$ and $1+\sigma-R$ ($1+\sigma-R$ is fixed rather
than $\sigma$ to avoid the complication of multiple solutions).


\end{appendix}


\begin{thebibliography}{}
%
\bibitem[Arakeri, Avila, Dada and Tovar(2000)]{Arakeri}
\textsc{Arakeri, J.H., Avila, F.E., Dada, J.M. \& Tovar, R.O.}
2000 Convection in a long vertical tube due to unstable
stratification- A new type of turbulent flow?
\emph{Current Science} \textbf{79}, 859-866.

\bibitem[Batchelor \& Nitsche (1993)]{Batchelor}
\textsc{Batchelor, G.K. \& Nitsche, J.M.}
1993 Instability of stratified fluid in a vertical cylinder.
\emph{J. Fluid Mech.} \textbf{252}, 419-448.

\bibitem[Beckett et al. (2009)]{Beckett}
\textsc{Beckett, F., Witham, F., Phillips, J.C. \& Mader, H.}
private communication concerning experiments curently being carried
out in the Department of Earth Sciences, University of Bristol -
preprint coming.

\bibitem[Charles and Redberger(1961)]{Charles}
\textsc{Charles, M.E. \& Redberger, R.J.}
1961 The reduction of pressure gradients in oil pipelines by the
addition of water. Numerical analysis of stratified flows
\emph{Can. J. Chem. Engng} \textbf{40}, 70-75.

\bibitem[Frigaard \& Scherzer(1998)]{Frigaard}
\textsc{Frigaard, I.A. \& Scherzer, O.}
1998 Uniaxial exchange flows of Bingham fluids in a cylindrical duct
\emph{IMA J. App. Math.} \textbf{61}, 237-266.

\bibitem[Hasson, Mann \& Nir(1970)]{Hasson}
\textsc{Hasson, D., Mann, U. \& Nir, A.}
1970 Annular flow of two immiscible liquids. I. Mechanisms.
\emph{Can. J. Chem. Engng.} \textbf{48}, 514.

\bibitem[Huppert \& Hallworth (2007)]{Huppert}
\textsc{Huppert, H.E. \& Hallworth, M.A.}
2007 Bi-directional flows in constrained systems
\emph{J. Fluid Mech.} \textbf{578}, 95-112.

\bibitem[Joseph, Renardy \& Renardy(1984)]{Joseph}
\textsc{Joseph, D.D., Renardy, M. \& Renardy, Y.}
1984 Instability of the flow of two immiscible liquids with
different viscosities in a pipe.
\emph{J. Fluid Mech.} \textbf{14}, 309-317.

\bibitem[Joseph, Nguyen and Beavers(1984)]{Joseph1}
\textsc{Joseph, D.D., Nguyen, K. \& Beavers, G.S.}
1984 Non-uniqueness and stability of the configuration of flow of
immiscible fluids with different viscosities.
\emph{J. Fluid Mech.} \textbf{14}, 319-345.

\bibitem[Joseph, Bai, Chen Renardy]{Joseph2}
\textsc{Joseph, D.D., Bai, R., Chen, K.P. \& Renardy, Y.Y.}
1997 Core-annular flows
\emph{Ann. Rev. Fluid Mech.} \textbf{29}, 65-90.

\bibitem[Lee and White(1974)]{Lee}
\textsc{Lee, B.L. \& White, J.L.}
1974 An experimentalstidy of rheological properties of polymer melts
in laminar shear flow and of interface defomration and its mechanisms
in two-phase stratified flow.
\emph{Trans. Soc. Rheol.} \textbf{18}, 467.

\bibitem[Maclean(1973)]{Maclean}
\textsc{Maclean, D.L.}
1973 A theoretical analysis of bicomponent flow and the problem of
interface shape
\emph{Trans. Soc. Rheol.} \textbf{17}, 385.

\bibitem[Markushevich(1965)]{Markushevich}
\textsc{Markuskevich, A. I.}
1965 Theory of Functions of a Complex Variable, vol 1
\emph{Prentice-Hall, Inc.} Englewood Cliffs, New Jersey (p205-207)

\bibitem[Minagawa and White(1975)]{Minagawa}
\textsc{Minagawa, N. \& White, J.L.}
1975 Coextrusion of unfilled and Ti0$_2$-filled polyethylene:
influence  of viscosity and die cross-section on interface shape.
\emph{Polymer Engng Sci.} \textbf{15}, 825.

\bibitem[Moyers-Gonzalez \& Frigaard(2004)]{Moyers}
\textsc{Moyers-Gonzalez, M.A. \& Frigaard, I.A.}
2004 Numerical solution of duct flows of multiple visco-plastic fluids
\emph{J. Non-Newtonian Fluid Mech.} \textbf{122}, 227-241.


\bibitem[Seon et al(2007)]{Seon}
\textsc{Seon, T, Znaien, J., Salin, D., Hulin, J.P., Hinch, E.J. \&
  Perrin, B.}
2007 Transient buoyany-driven front dynamics in nearly horizontal tubes
\emph{Phys. Fluids} \textbf{19}, 123603

\bibitem[Southern and Ballman(1973)]{Southern}
\textsc{Southern, J.H. \& Ballman, R.L.}
1973 Stratified bicomponent flow of polymer melts in a tube
\emph{Appl. Polymer Symp.} \textbf{20}, 175-189.

\bibitem[Taghavi(2009)]{Taghavi}
\textsc{Taghavi, S.M., Seon, T., Martinez, D.M. \& Frigaard, I.A.}
2009 Buoyancy-dominated displacement flows in near-horizontal
channels: the viscous limit
\emph{J. Fluid Mech.} \textbf{639}, 1-35.

\bibitem[Vlasov(1986)]{Vlasov}
\textsc{Vlasov, V.I.}
1986 Solution of a Dirichlet problem in a crescent-shaped domain
\emph{J. Eng. Phys. \& Thermophys.} \textbf{50}, 741-747.


\bibitem[White(1991)]{White}
\textsc{White, F. M.}
Viscous Fluid Flow
\emph{McGraw-Hill} (p124)

\bibitem[Williams(1975)]{Williams}
\textsc{Williams, M.C.}
1975 Migration of two liquid phases in capillary extrusion: an energy interpretation
\emph{AIChE. J.} \textbf{21}, 1204.

\bibitem[Yu and Sparrow(1967)]{Yu}
\textsc{Yu, H.S. \& Sparrow, E.M.}
1967 Straified laminar flow in ducts of arbitrary shape
\emph{AIChE. J.} \textbf{13}, 10.

\bibitem[Znaien et al(2009)]{Znaien} 
\textsc{Znaien, J., Hallez, Y., Moisy, F., Magnaudet, J., Hullin,
  J.P., Salin, D. \& Hinch, E.J.}  
2009 Experimental and numerical investigations of
flow structure and momentum transport in a turbulent buoyancy-driven
flow inside a tilted tube.  \emph{Phys. Fluids} \textbf{21}, 115102.

\end{thebibliography}
\end{document}